\documentclass[a4paper,11pt]{article}
\pdfoutput=1
\usepackage{jheppub}
\usepackage{amsmath,amssymb,bm,graphicx,epsf,colordvi}
\usepackage{slashed}
\def\bea#1\eea{\begin{align}#1\end{align}} 
\usepackage{diagbox}
\usepackage{mathtools}
\usepackage{lipsum}
\usepackage{soul}
\usepackage{enumitem}
\usepackage{braket}
\usepackage{bm}
\allowdisplaybreaks 
\usepackage{color}
\usepackage[dvipsnames]{xcolor}

\usepackage{mathrsfs}
\usepackage{appendix}
\usepackage{bm}
\usepackage{lipsum}

\newcommand{\bef}{\begin{figure}[hbt]\centering}
\newcommand{\eef}{\end{figure}}
\definecolor{darkgreen}{rgb}{0.13,0.55,0.13}
\definecolor{darkpurple}{rgb}{0.5, 0.2, 0.8}
\definecolor{or}{rgb}{0.88,0.43,0.02}
\usepackage{mathtools}
\usepackage{subfigure}
\usepackage{subcaption}
\usepackage{booktabs}
\usepackage{graphicx}
\graphicspath{ {./figure/} }
\usepackage{caption}
\usepackage{lipsum}

\def\cW{{\mathcal W}}
\def\cV{{\mathcal V}}

\newcommand{\beq}{\begin{equation}}
\newcommand{\eeq}{\end{equation}}
\def\bea#1\eea{\begin{align}#1\end{align}}

\newcommand{\f}{\frac}
\def \be  {\begin{equation}}
\def \ee  {\end{equation}}
\def \ba  {\begin{eqnarray}}
\def \ea  {\end{eqnarray}}
\newcommand{\df}{\mathrm{d}}

\newcommand{\nn}{\nonumber}
\newcommand\as{\alpha_s}

\newcommand{\sij}{\tilde{s}_{ij}}
\newcommand{\sik}{\tilde{s}_{ik}}
\newcommand{\sjk}{\tilde{s}_{jk}}
\newcommand{\zjk}{\left(\frac{x_j}{x_k}\right)^{2\alpha}}

\allowdisplaybreaks

\usepackage{graphicx}

\DeclareRobustCommand{\Sec}[1]{sec.~\ref{#1}}
\DeclareRobustCommand{\App}[1]{App.~\ref{#1}}
\DeclareRobustCommand{\Eq}[1]{eq.~(\ref{#1})}
\newcommand{\eq}[1]{Eq.~\eqref{eq:#1}}

\usepackage{stackengine}

\title{Revisiting Single Inclusive Jet Production:\\ Timelike Factorization and Reciprocity}
\preprint{MIT-CTP 5766}
\author[a]{Kyle Lee,}
\affiliation[a]{Center for Theoretical Physics, Massachusetts Institute of Technology, Cambridge, MA 02139}

\author[b]{Ian Moult,}
\affiliation[b]{Department of Physics, Yale University, New Haven, CT 06511}

\author[c]{Xiaoyuan Zhang}
\affiliation[c]{Department of Physics, Harvard University, Cambridge, MA 02138}

\emailAdd{kylel@mit.edu,ian.moult@yale.edu,xiaoyuanzhang@g.harvard.edu}

\abstract{Factorization theorems for single inclusive jet production play a crucial role in the study of jets and their substructure.
In the case of small radius jets, the dynamics of the jet clustering can be factorized from both the hard production dynamics, and the dynamics of the low scale jet substructure measurement, and is described by a matching coefficient that can be computed in perturbative Quantum Chromodynamics (QCD).
A proposed factorization formula describing this process has been previously presented in the literature, and is referred to as the semi-inclusive, or fragmenting jets formalism.
By performing an explicit two-loop calculation, we show the inconsistency of this factorization formula, in agreement with another recent result in the literature.
Building on recent progress in the factorization of single logarithmic observables, and the understanding of reciprocity, we then derive a new all-order factorization theorem for inclusive jet production. 
The use of a jet algorithm, being only a modification of the infrared structure of the measurement, modifies the structure of convolutions in the factorization theorem, as compared to inclusive fragmentation, but maintains the universality of the inclusive hard function and its associated Dokshitzer-Gribov-Lipatov-Altarelli-Parisi (DGLAP) evolution, which are ultraviolet properties. 
However, the non-trivial structure of convolutions in the factorization theorem implies that the jet functions exhibit a modified evolution.
We perform an explicit two-loop calculation of the jet function in both $\mathcal{N}=4$ super Yang-Mills (SYM), and for all color channels in QCD, finding exact agreement with the structure derived from our renormalization group equations.
In addition, we derive several new results, including an extension of our factorization formula to jet substructure observables, a jet algorithm definition of a generating function for the energy correlators, and new results for exclusive jet functions.
Our results are a key ingredient for achieving precision jet substructure at colliders.
}

\begin{document}

\maketitle

\section{Introduction}
The inclusive production of energetic jets is one of the most fundamental processes in collider physics, providing myriad applications—from searching for new physics to precision measurements of Standard Model parameters to probing nuclear structure. Furthermore, a precise description of the inclusive jet production is crucial for the jet substructure program~\cite{Larkoski:2017jix,Kogler:2018hem,Marzani:2019hun}, where additional measurements are performed on the produced jet. Among different jet production mechanisms, single inclusive production of jets in various collider environments, such as $pp\to \text{jet}X$, $e^+e^-\to \text{jet}X$, and $ep\to \text{jet}X$, offer the highest statistics of jet cross section by measuring all jets within specific transverse momentum $p_T$ and rapidity intervals without further restrictions. As a result, single inclusive jet process has been computed and measured across various collider environments, including $pp$, $ep$, and $e^+e^-$ colliders~\cite{ZEUS:2012pcn,H1:2016goa,CMS:2020caw,ALICE:2019qyj,ATLAS:2017ble,ATLAS:2017kux,CMS:2016jip,Currie:2018xkj,Czakon:2019tmo}. For a review, see \cite{Gehrmann:2021qex}. 

Jets are identified using a jet algorithm, primarily the anti-$k_T$ algorithm \cite{Cacciari:2005hq,Salam:2007xv,Cacciari:2008gp,Cacciari:2011ma}, which is defined by a jet radius, $R$. For the single inclusive jet process, this introduces a hierarchy between two scales, $p_T$ and $p_T\,R$ (or $Q$ and $Q\,R/2$ for measuring jet energy in $e^+e^-$ with $\sqrt{s}=Q$) for small jet radius. Factorizing the dynamics at these scales offers several advantages~\cite{Kang:2016mcy,Dai:2016hzf}. First, it enables the resummation of logarithms of $\alpha_s^n \ln^n R$ between the jet scale and the hard scale, improving the perturbative stability of the result. Second, it simplifies the calculation significantly, as the hard function for jet production matches the inclusive hard function of hadron fragmentation. The jet function, which encapsulates all aspects of the clustering measurement at the perturbative scale $p_T\,R$,  is then completely factorized from the universal inclusive hard function. This factorization is particularly useful when additional measurements of jet substructure observables are performed on the produced jet, as the factorization between the hard and jet scales is retained. Therefore, one only needs to modify the jet functions for different jet substructure observables, which can be refactorized to separate the production of the jets at the scale $p_T\,R$ and some jet substructure scale at $\mu_v$ for a jet substructure $v$. This powerful factorization structure has facilitated numerous studies on jet substructure observables with single inclusive production of jets~\cite{Hannesdottir:2022rsl,Lee:2023xzv,Craft:2022kdo,Chien:2018lmv,Lee:2022ige,Lee:2023npz,Aschenauer:2019uex,Kang:2018qra,Cal:2020flh,Kang:2020xyq,Kang:2018vgn,Kang:2018jwa,Kang:2019prh,Cal:2021fla,Cal:2019gxa,Kang:2016ehg,Cal:2019hjc,Kang:2017mda,Mehtar-Tani:2024smp}.

Among different jet substructure observables, energy correlator observables stand out as the most precisely studied observable with considerable recent interest and wide applications~\cite{Bossi:2024qho,Holguin:2024tkz,Holguin:2023bjf,Xiao:2024rol,Holguin:2022epo,Chen:2023zlx,Lee:2022ige,Chen:2020vvp,Jaarsma:2022kdd,Jaarsma:2023ell,Chen:2022muj,Lee:2023npz,Lee:2023tkr,Lee:2024esz,Chen:2024nyc,Andres:2023xwr,Andres:2022ovj,Devereaux:2023vjz,Andres:2023ymw,Andres:2024ksi,Barata:2024nqo,Barata:2023bhh,Cao:2023oef,Chen:2024bpj,Liu:2023aqb,Liu:2022wop,Craft:2022kdo,Chen:2020adz,Yan:2022cye,Yang:2022tgm,Yang:2024gcn,Chang:2022ryc,Chen:2019bpb}. The jet function describing the measurement of the projected energy correlators at two \cite{Dixon:2019uzg}, and three-points is now known at two loops \cite{Chen:2023zlx}. These observables have been measured inside high energy jets \cite{Komiske:2022enw,CMS:2024mlf,Tamis:2023guc}, and are used by CMS to extract the value of $\alpha_s$ at $4\%$ accuracy~\cite{CMS:2024mlf}. The dominant theoretical uncertainty in this result is the perturbative accuracy, highlighting the importance of improving the accuracy of the different perturbative components involved.

While substantial progress has been made in computing many of the necessary perturbative components for the jet substructure program at higher perturbative orders—such as hard functions~\cite{Currie:2018xkj,Czakon:2019tmo,Czakon:2021ohs,Goyal:2024tmo,Goyal:2023zdi,Bonino:2024wgg,Bonino:2024qbh} and jet functions describing the measurement of observables like energy correlators—the inclusion of clustering contributions for jet production has lagged behind and is fully understood only at one loop~\cite{Kang:2016mcy}, with an important exception of the exclusive quark jet contribution at two loops from~\cite{Liu:2021xzi}. As such, they are one of the leading perturbative uncertainties for processes involving identified jet production, as illustrated in a recent study of $H+1$ jet production~\cite{Cal:2024yjz}, for example. From a broader perspective, these functions provide the crucial matching between hard scales and infrared jet measurements in jet substructure studies. They are the only remaining barrier to interfacing two-loop calculations of jet substructure observables with state-of-the-art hard scattering amplitude calculations. Therefore, improving our understanding of these functions, along with performing their explicit perturbative calculations, is essential for advancing the perturbative accuracy of the jet substructure program.

Here we initiate a series of papers to improve upon the situation. In this paper, we begin by re-examining the factorization theorem for the single inclusive production of jets. Building on recent progress in factorization theorems for energy correlators, we present an all-orders factorization theorem for single inclusive jet production that differs from previously established formulations. Our factorization theorem incorporates the universal hard function from hadron fragmentation and features renormalization group evolution determined entirely by the universal timelike DGLAP splitting kernels. However, due to a non-trivial modification in the convolution structure, which results from the identification of a jet state rather than a single hadron state, the resulting RG evolution is not equivalent to timelike DGLAP. Despite this difference, the connection to timelike anomalous dimensions allow us to extend the accuracy of this process to next-to-next-to-leading logarithms (NNLL) immediately. We perform an explicit two-loop calculation of the jet function, finding perfect agreement with the prediction from an RG analysis. We believe that our factorization theorem will be central to improving the perturbative description of jets at collider experiments.

Due to the fundamental nature of this process, a number of different approaches have been presented to describe it, so we briefly comment on the relation to our approach. In \cite{Dasgupta:2014yra,Dasgupta:2016bnd}, a generating functional approach was used. This approach has been extended to next-to-leading logarithms (NLL) in \cite{vanBeekveld:2024jnx,vanBeekveld:2024qxs}, and implemented in a collinear parton shower algorithm to achieve NLL resummation for fragmentation observables. In SCET \cite{Bauer:2000ew,Bauer:2000yr,Bauer:2001ct,Bauer:2001yt,Rothstein:2016bsq}, a factorization formula was presented in \cite{Kang:2016mcy}, which introduced a ``semi-inclusive jet function". This was later extended to involve an identified momentum fraction inside the jet, and is termed ``fragmenting jet function" \cite{Kang:2016ehg,Kang:2020xyq}. Assuming the structure of their factorization formulas, they were able to show using renormalization group consistency arguments, that the evolution between jet scale $p_T\, R$ and the hard scale $p_T$ are governed by the usual timelike DGLAP evolution for hadron fragmentation process. We will show by explicit calculation that this is incorrect, thus rendering the factorization formula presented in \cite{Kang:2016mcy,Kang:2016ehg} inconsistent. This conclusion was also found in \cite{vanBeekveld:2024jnx}. While we agree with the results of the calculation of \cite{vanBeekveld:2024jnx}, we are able to interpret this not as a modification of the anomalous dimension, but rather as a modification of the factorization theorem. This observation will allow us to present a new all-order factorization for the single inclusive jet process.

Our factorization theorem has a number of advantages. Most importantly, it maintains the universality of the hard function, and its RG evolution is entirely determined by the timelike DGLAP anomalous dimensions and their derivatives. Using known results for the timelike anomalous dimensions~\cite{Mitov:2006ic,Mitov:2006wy,Chen:2020uvt}, this means that we can immediately extend our formalism to NNLL.

An outline of this paper is as follows. In \Sec{sec:fact} we define our observable and present an all-orders factorization theorem for single inclusive jet production with a small jet radius. We explore its relation to reciprocity and its connection to factorization theorems for energy correlators, while also commenting on relations to previous factorization formulas in the literature. In \Sec{sec:fact_gen}, we present numerous generalizations of our factorization theorem relevant to jet substructure studies. In \Sec{sec:EECalg}, we leverage the intimate connection between the factorization structure of energy correlators and single inclusive jet production to develop a new jet algorithm directly associated with the energy correlator observables. In \Sec{sec:renorm} we carry out the renormalization group analysis of the jet function to derive the logarithmic structure and associated poles at two loops. In \Sec{sec:calc} we describe the two-loop calculation of the jet function, and present the numerical results of the anomalous dimensions for $\mathcal{N}=4$ super Yang-Mills and for all color channels in QCD, finding perfect agreement with the predictions from our renormalization group analysis. We conclude in \Sec{sec:conc}. 

\textbf{Note:} For the remainder of the paper, we will often simply refer to single inclusive hadron production as hadron fragmentation and to single inclusive jet production as inclusive jet production for convenience.

\section{Factorization Theorem for Small Radius Inclusive Jet Production}\label{sec:fact}
In this section we present an all-order factorization theorem for single inclusive small radius jet production, which separates the dynamics at the hard scale $Q$ from those at the jet scale $Q\,R/2$. To situate this factorization theorem, we begin by reviewing the factorization theorems for single inclusive hadron fragmentation and energy correlators, while also exploring their relationship to reciprocity. These discussions highlight notions of timelike factorization that are crucial for the factorization theorem for single inclusive jet production. We then present the factorization theorem for the single inclusive jet production, discuss its structure in detail, and compare it with other approaches in the literature.

\subsection{Review of Hadron Fragmentation, Energy Correlators, and Reciprocity}\label{sec:fact_review}

We begin by reviewing factorization theorems for hadron fragmentation and the energy correlators. These highlight, in a controlled and calculable setting, how the modification of the IR measurement modifies the anomalous dimension.

The factorization theorem for inclusive hadron fragmentation has been proven rigorously \cite{Collins:1981ta,Bodwin:1984hc,Collins:1985ue,Collins:1988ig,Collins:1989gx,Collins:2011zzd,Nayak:2005rt}, and applied in $e^+e^-$, $ep$ and $pp$ collisions. For high-$p_T$ hadron production process, this is given as
\begin{align}
\label{eq:highpTfrag}
\frac{d\sigma_h}{dp_T d\eta}&= \int_0^{\infty} \mathrm{d} p_T^{\prime} \int_{z^{\rm min}}^1 \mathrm{~d} z\, \vec{\mathcal{H}}\left(p_T^{\prime}, \eta, \mu\right)\cdot \vec{D}_{h}\left(z, \mu\right) \delta\left(p_T-z p_T^{\prime}\right)\nn\\
&=\int_{z^{\rm min}}^1 \frac{\mathrm{d} z}{z} \vec{\mathcal{H}}\left(p_T / z, \eta, \mu\right)\cdot \vec{D}_{h}\left(z, \mu\right)\,,
\end{align}
where $p_T' = p_T/z$ is the transverse momentum of the parton initiating the high-$p_T$ hadron. As observed $p_T$ of the hadron does not give access to the partonic transverse momentum, this naturally gives rise to the convolution structure. Allowed range of $p_T'$ is determined by the observed hadron $p_T, \eta$, as well as $\sqrt{s}$, giving the lower bound $z^{\rm min} = 2p_T\cosh\eta/\sqrt{s}$. Such high transverse momentum factorization is natural for $pp$, where we need access to boost invariant quantity. On the other hand, if we are able to tag on the initial scale $Q$, as in the case of $e^+e^-$ with $\sqrt{s}=Q$, the factorization for observing energy fraction of the hadron $z_h = 2E_h/Q$ can be formulated as 
\begin{align}
\label{eq:frag}
\frac{d\sigma_h}{d z_h}&= \int dx\,dz~ \vec{H}\left(x,\frac{Q^2}{\mu^2},\mu \right) \cdot\vec{D}_{h} \left(  z,\mu \right) \delta(z_h -x z)\nn\\
&= \int_{z_h}^1 \frac{dx}{x}~ \vec{H}\left(x,\frac{Q^2}{\mu^2},\mu \right) \cdot\vec{D}_{h} \left(\frac{z_h}{x},\mu \right)\,,
\end{align}
The functions $\vec{D}_{h}$ are universal hadron fragmentation functions~\cite{Collins:1981uw,Collins:1989gx,Metz:2016swz}. Here, we use vector notation to indicate that both are vectors in the flavor space. For concreteness, we will discuss factorization theorems with hadron energy fraction, but generalization to high-$p_T$ production case follows immediately.

Independent of the nature of the collision, the hard function satisfies a universal renormalization group equation, which is the timelike DGLAP equation \cite{Gribov:1972ri,Dokshitzer:1977sg,Altarelli:1977zs}. Explicitly, this equation reads
\begin{align}
  \label{eq:hardRG}
  \frac{d \vec{H} (x, \frac{Q^2}{\mu^2},\mu)}{d \ln \mu^2} = - \int_x^1 \frac{dy}{y} \widehat P_T(y) \cdot \vec{H}\left(\frac{x}{y}, \frac{Q^2}{\mu^2},\mu\right) \,,
\end{align}
where $\widehat P_T$ is the singlet timelike splitting kernel matrix, which is known to three-loops~ \cite{Mitov:2006ic,Mitov:2006wy,Moch:2007tx,Chen:2020uvt}. They are given as
\begin{align}
  \label{eq:splitK}
  \widehat P_T = 
  \begin{pmatrix}
    P_{qq} &\hspace{0.15cm}  P_{qg}
\\
    P_{gq} &\hspace{0.15cm} P_{gg}
  \end{pmatrix} \,.
\end{align}
On the other hand, the RG consistency and the convolution structure in \eq{frag} imply that $\vec{D}_{h}$ evolves as
\begin{align}
  \label{eq:hadRG}
  \frac{d \vec{D}_h (z,\mu)}{d \ln \mu^2} = \int_z^1 \frac{dy}{y}  \vec{D}_h\left(\frac{z}{y},\mu\right) \cdot \widehat P_T(y)\,,
\end{align}
which is also referred to as the (timelike) DGLAP evolution equation for hadron fragmentation functions.

If we now take $N$-th moments in $z_h$ for $d\sigma_h/dz_h$ of \eq{frag}, the Mellin convolution structure simplifies into a product form as
\begin{align}
\label{eq:hadmom}
\int_0^1 dz_h\,z_h^{N}\,\frac{d\sigma_h}{dz_h} = \vec{H}(N,\frac{Q^2}{\mu^2},\mu)\cdot \vec{D}_h (N,\mu)\,,
\end{align}
where 
\begin{align}
\label{eq:hadDGmom}
\vec{H}(N,\frac{Q^2}{\mu^2},\mu) &\equiv \int_0^1 dx\, x^{N}\,\vec{H} (x,\frac{Q^2}{\mu^2},\mu)\,,\nn\\
\vec{D}_h (N,\mu) &\equiv \int_0^1 dz\, z^{N}\,\vec{D}_h (z,\mu)\,.
\end{align}
Now the RG evolution is governed by the \emph{timelike} DGLAP anomalous dimensions $\gamma_T(N)$ as
\begin{align}
\label{eq:hadmomRG}
\frac{d \vec{H}(N,\frac{Q^2}{\mu^2},\mu)}{d \ln \mu^2} &= \gamma_T(N+1)\cdot \vec{H}(N,\frac{Q^2}{\mu^2},\mu)\,,\nn\\
\frac{d \vec{D}_h (N,\mu)}{d \ln \mu^2} &= -\vec{D}_h (N,\mu) \cdot \gamma_T(N+1)\,.
\end{align}
We will later contrast how these RG structures are modified for the case involving jet production. The timelike DGLAP anomalous dimensions are defined as
\begin{equation}
\label{eq:gammaPTrelation}
\gamma_T(k)\ \equiv\ - \int_0^1 dy \, y^{k-1} \, \widehat{P}_T(y)\,,
\end{equation}
where the $k$ for the \emph{spacelike} anomalous dimensions counterpart represent the \emph{spin} of the twist-2 local operator. 

Before turning to the case of identified jet production, we first consider a case of intermediate complexity, where the infrared measurement is modified to give rise to an IR safe observable. This is the case of the energy correlators \cite{Basham:1979gh,Basham:1978zq,Basham:1978bw,Basham:1977iq,Hofman:2008ar} in the small angle limit. The small angle limit factorization is understood from a number of different perspectives, \cite{Korchemsky:2019nzm,Dixon:2019uzg,Kologlu:2019mfz,Chen:2023zzh}. An important aspect of the factorization theorems is that they can be tested against explicit analytic calculations of the observable, which are available at next-to-leading order (NLO) in QCD \cite{Dixon:2018qgp} and NNLO in $\mathcal{N}=4$ super Yang-Mills theory \cite{Belitsky:2013ofa,Henn:2019gkr}. These calculations provide valuable insight into the structure of collinear limits in QCD, which can then be generalized using factorization theorems, to apply to much more general settings in collider physics, and have a large phenomenological impact.

General $N$-point energy correlators of the form $\left\langle\Psi\left|\mathcal{E}\left(\vec{n}_1\right) \mathcal{E}\left(\vec{n}_2\right) \cdots \mathcal{E}\left(\vec{n}_N\right)\right| \Psi\right\rangle$ measure the distribution of energy in some state $|\Psi\rangle$ created by high energy collisions. The energy flow pattern is captured by the energy flow operator
\begin{align}
\mathcal{E}(\vec{n})=\int_0^{\infty} d t \lim _{r \rightarrow \infty} r^2 n^i T_{0 i}(t, r \vec{n})\,,
\end{align}
which relates the energy-momentum tensor $T_{\mu\nu}$ in QCD to the energy captured by a detector in the direction $\vec{n}$. In particular, the relevant projection of the general $N$-point energy correlators is the \emph{projected} $N$-point energy correlators~\cite{Chen:2020vvp}, which measure only the largest angle $x_L \equiv\left(1-\cos \theta_L\right) / 2=\left(1-\text{min}(\vec{n}_i \cdot \vec{n}_j)\right) / 2$. In the small angle limit of the projected $N$-point energy correlators, the cumulant is defined as
\begin{align}
\Sigma^{[N]}\left(x_L\right)\equiv\frac{1}{\sigma} \int_0^{x_L} d x_L^{\prime} \frac{d \sigma^{[N]}}{d x_L^{\prime}}\,,
\end{align}
exhibits factorization structure derived from SCET as
\begin{align}
\label{eq:fact_N}
 \Sigma^{[N]}(x_L)
= \int_0^1 dx\, x^N \vec{J}^{[N]} (\ln\frac{x_L x^2 Q^2}{\mu^2},\mu)
   \cdot  \vec{H} (x,\frac{Q^2}{\mu^2},\mu) \,.
\end{align}
Here $H$ is the inclusive hard function, identical to that for the case of fragmentation in \eq{frag}, and thus obeys the identical RG equations of the timelike DGLAP in \eq{hardRG}. As compared to the case of moments of hadron fragmentation presented in \eq{hadmom}, since this is only a modification of the IR measurement, it merely modifies the convolution structure, and the structure of the jet function $J^{[N]}$.  

By RG consistency, the RG evolution of the jet function is fully determined by the timelike splitting kernels as the hard function. Specifically, it is given by
\begin{align}
  \label{eq:jetRG}
\frac{d \vec{J}^{[N]}(\ln\frac{x_L Q^2}{\mu^2}, \mu) }{d \ln \mu^2} = \int_0^1 dy\, y^N \vec{J}^{[N]} (\ln\frac{x_L y^2 Q^2}{\mu^2}, \mu) \cdot \widehat P_T(y,\mu) \,.
\end{align}
This differs from the standard DGLAP equation for hadron fragmentation functions in \eq{hadDGmom}. For instance, only by ignoring the convolution between the jet function and the timelike splitting kernel matrix (setting $y^2\to 1$ in the jet function argument) would we recover the standard DGLAP equation, governed by the timelike anomalous dimension, as the $N$-th moment of the hadron fragmentation function in \eq{hadmomRG}. The new convolution structure introduces a renormalization group evolution governed not only by the timelike anomalous dimensions but also by their derivatives. This subtle change in the convolution structure of the factorization theorem leads to a significant modification of the power-law scaling of energy correlators. Indeed, we will find that in a CFT, it is not the timelike, but rather the \emph{spacelike} anomalous dimension that controls its behavior! To highlight this fact, we repeat the analysis of \cite{Dixon:2019uzg}, which will be identical to the case of the moments of inclusive jet production as we will see later. However, in the case of the energy correlators, all the fixed-order calculations can be performed analytically to verify the results.

Since the coupling does not run in a CFT, we can take a power law ansatz for the resulting jet function as 
\begin{align}
  \label{eq:Jneqfour}
  J^{[N]}(x_L\,Q^2,\mu) = C^{[N]}_J(\alpha_s) \left(\frac{x_L\,Q^2}{\mu^2} \right)^{\gamma_{J,[N]}^{{\cal N}=4}(\alpha_s)} \,,
\end{align}
where we consider $\mathcal{N}=4$ super-Yang-Mills theory case for concreteness. Substituting this to~\eq{fact_N} gives rise to the scaling behavior of the energy correlator observable as
\begin{equation}
\label{eq:CFTscaling}
\Sigma^{[N]}(x_L) = \frac{1}{2} \, C(\alpha_s) \, x_L^{\gamma_{J,[N]}^{{\cal N}=4}(\alpha_s)} \,.
\end{equation}
On the other hand, substituting our jet function ansatz into the RG evolution equation for the jet function given in \eq{jetRG} and the definition of timelike anomalous dimension given in \eq{gammaPTrelation}, we find\footnote{Note that we use the shifted argument for $\mathcal{N}=4$ due the supermultiplet sum~\cite{Kotikov:2004er}.}
\begin{align}
  \label{eq:gammaNeqfour}
\gamma_{J,[N]}^{{\cal N}=4}(\alpha_s) = &\, - \int_0^1 dy\, y^{N + 2 \gamma_J^{{\cal N}=4}(\alpha_s)} P_{T, \text{uni.}}(y,\alpha_s)
\nn\\
=&\, \gamma_T^{{\cal N}=4}(N-1 + 2 \gamma_J^{{\cal N}=4}, \alpha_s) \,,
\end{align}
Comparing this with the reciprocity relation~\cite{Mueller:1983js,Dokshitzer:2005bf,Marchesini:2006ax,Basso:2006nk,Dokshitzer:2006nm,Caron-Huot:2022eqs}, which relates the spacelike and timelike anomalous dimensions (In the case that there are multiple anomalous dimensions, it applies to their eigenvalues),
\begin{align}\label{eq:reciprocity}
2 \gamma_S^{{\cal N}=4}(N , \alpha_s)=  2 \gamma_T^{{\cal N}=4}(N + 2 \gamma_S^{{\cal N}=4}, \alpha_s), 
\end{align}
we find
\begin{align}
 \gamma_{J,[N]}^{{\cal N}=4}(\alpha_s) = \gamma_S^{{\cal N}=4}(N-1,\alpha_s).
 \label{eq:anomN4}
 \end{align}
In other words, the nontrivial convolution structure in the factorization of energy correlators causes the power-law scaling in \eq{CFTscaling} to be governed by the \emph{spacelike} anomalous dimension, rather than the timelike anomalous dimension that the factorization theorem is expressed in terms of. 

\begin{figure}
\begin{center}
\includegraphics[scale=0.75]{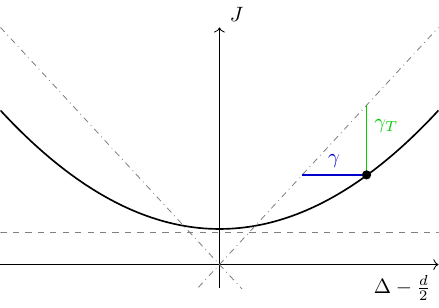} 
\end{center}
\caption{The Regge trajectories illustrate the light-ray operators in both free and interacting theories. A given operator on the interacting trajectory can be viewed as horizontally or vertically displaced from one on the free trajectory, with these displacements corresponding to spacelike and timelike anomalous dimensions, respectively. Since the free trajectory forms a 45-degree angle, the two displacements are related by what is known as the reciprocity relation.}
\label{fig:reciprocity}
\end{figure} 

From a modern perspective \cite{Caron-Huot:2022eqs}, the reason for this is clear: the timelike and spacelike anomalous dimensions are simply different choices for specifying a single operator in the theory, and in SCET factorization theorems in \eq{fact_N}, we have chosen to work in terms of the timelike anomalous dimension. To understand this,  we recall that the dimensions of twist-two operators are analytic in spin \cite{Caron-Huot:2017vep}, and lie on Regge trajectories shown in  Fig.~\ref{fig:reciprocity} defined by $\Delta=\Delta (J)$, or inverting this relation $J=J(\Delta)$ \cite{Kravchuk:2018htv}, where $\Delta$ and $J$ denote the dimension and spin, respectively. The operators that live on these trajectories are light-ray operators~\cite{Kravchuk:2018htv}, where at integer values of $J$, they are formed by light-transform of local operators.
The linear trajectory in the dashed line corresponds to the twist-2  operators in the free theory and its shadow, while the curved trajectory corresponds to operators in the interacting theory.  Given a particular operator on the interacting trajectory, there are two ways of specifying its location: We can either give a fixed value of $J$, and then compute the spacelike anomalous dimension $\gamma(J)$, which is the horizontal shift between the free and interacting trajectories. 
This is most common for operators that have a fixed $J$, as for the case involving local operators. Alternatively, we can fix the value of $\Delta$, and then specify the vertical displacement $\gamma_T(\Delta-\frac{d}{2})$, which corresponds to the timelike anomalous dimension. This is what is done in the case of fragmentation.  Generically in this case one obtains an operator with a non-integer spin, which corresponds to a light-ray operator, which is not defined in terms of a local operator for non-integer spin. This is why it is commonly stated that the timelike anomalous dimension controlling fragmentation is not the anomalous dimension of a local operator.

From the fact that the twist-two trajectory in the free theory is a $45$ degree line, we can immediately read off the geometric relation between the two anomalous dimensions
\begin{align}
\gamma_T(N)=\gamma(N-\gamma_T(N))\,,
\end{align}
providing a simple geometric interpretation of reciprocity. Namely, it is just whether one considers the anomalous shift in $\Delta$ or $J$ to specify the same operator. These two ways of specifying the operators on the trajectory correspond to traditional conformal frame and detector frame in a CFT~\cite{Caron-Huot:2022eqs}.

The formulation of the factorization in SCET for timelike jet observables is in terms of the timelike anomalous dimension. However, this is merely a choice of organization. Indeed, using the light-ray OPE to describe the small angle limit of the energy correlators, one naturally works entirely in terms of the spacelike anomalous dimensions \cite{Hofman:2008ar,Kologlu:2019mfz}. However, using the timelike anomalous dimension to formulate the factorization does not imply that SCET factorization theorems cannot describe observables that scale with the spacelike anomalous dimension. In SCET this is incorporated through the specific structure of the convolution, which modifies the scaling to be governed by spacelike anomalous dimension as discussed around~\eq{anomN4}.

We choose to emphasize this point, since we will see that \emph{exactly} the same phenomenon occurs in the case of inclusive jet production. It was recently found that \cite{vanBeekveld:2024jnx} that the timelike scaling predicted by the factorization formula of \cite{Kang:2016mcy} is incorrect. While we agree with the observation of \cite{vanBeekveld:2024jnx}, we stress that this discrepancy arises from the convolution structure in the factorization formula of \cite{Kang:2016mcy}, not from the anomalous dimension itself. That is, the dependence on the derivatives of the timelike anomalous dimension results from changes in the convolution structure due to modifications in the IR measurement, rather than changes in the anomalous dimensions, which are inherently UV in nature. The example of the energy correlator highlights how, if the convolution structure is correct, factorization formulas involving objects, such as the inclusive hard function, that obey timelike DGLAP, can easily give rise to final results that obey spacelike DGLAP. Of course, the object of physical interest, is the physical behavior of the final observable, not how one chooses to specify the operators used to describe this behavior.

We can also extend this analysis beyond the conformal case. This analysis was performed for the pure Yang-Mills case in \cite{Dixon:2019uzg} for two-point energy correlators. By making the ansatz
\begin{align}\label{eq:ansatz_YM}
&J_g^{[N]}\left(\frac{\mu^2}{zQ^2}, \alpha_s(\mu) \right)  =\ C_J (\alpha_s(\sqrt{z}Q)) \exp \Biggl[
 - \! \!\!\int\limits_{\alpha_s(\sqrt{z}Q)}^{\alpha_s(\mu)}   \!\!\! d\bar \alpha_s \frac{\gamma^{\text{YM}}_{J,[N]}(\bar \alpha_s, z)}{\beta(\bar \alpha_s)} \Biggr] \,, 
\end{align}
and repeating the exercise in~\cite{Dixon:2019uzg}, the effective anomalous dimension is given to NLL for projected $N$-point energy correlators as
\begin{align}
\label{eq:pureYMNLL}
\gamma_{J,[N]}^{\text{YM}}&=\gamma^{\text{YM}}_S(N+1)  - \gamma^{\text{YM}}_S(N+1) \partial_M \gamma^{\text{YM}}_S(M)\Big|_{M=N+1}
\frac{\alpha_s(Q)}{4\pi} 2 \beta_0 \ln z +\cdots\,. 
\end{align}
Therefore, in non-CFT, the effective anomalous dimension is not only given by the spin-$N+1$ spacelike anomalous dimensions, but its derivative $\partial_M \gamma_S^{\rm YM}(M)$ with coefficients proportional to the beta function, which can be seen as ``tuning" the anomalous dimension in the final physical result.

This procedure can be iterated to any logarithmic order. We wish to emphasize that this is not a real anomalous dimension, but rather an effective anomalous dimension.  This is similar in spirit to what was used by \cite{vanBeekveld:2024jnx} in the case of inclusive jet production, as will be discussed in more detail later. The advantage of formulating the factorization in terms of field theory operators, is that they can be extracted to high perturbative orders from independent calculations, and the effective anomalous dimensions of the final observables are then fully determined in terms of these few universal anomalous dimensions. We will demonstrate how the derivatives of timelike anomalous dimension naturally make appearance for full QCD case in sec.~\ref{sec:renorm} through modified convolution structure.

\subsection{Factorization Theorem for Inclusive Jet Prodution}\label{sec:fact_new}
We would now like to extend this factorization theorem from an inclusive identified hadron, to an inclusive small $R$ jet, defined with a jet algorithm, which we will take to be recombination jet algorithm of $k_T$-type \cite{Cacciari:2005hq,Salam:2007xv,Cacciari:2008gp,Cacciari:2011ma}, which includes anti-$k_T$ algorithm as an instance. Compared to hadron fragmentation, the use of the IR-safe jet algorithm makes the inclusive jet production IR safe. From the discussion of the previous section, since using a jet algorithm instead of an identified hadron is only a modification of the IR measurement,  we anticipate the following properties of the factorization theorem: 
\begin{enumerate}
	\item The factorization should not modify the hard function as compared to hadron fragmentation.
	\item The factorization does not modify that the renormalization group properties will be entirely determined by the timelike splitting kernels.
\end{enumerate}
This is advantageous, since it will enable us to perform high-order resummation in the identified jet case, since the timelike splitting kernels are already known. However, crucially, the modification of the infrared measurement, much like when switching between fragmentation and the energy correlator, \emph{can} modify the convolution structure between the hard function and the jet function in the factorization theorem, leading to a different behavior of the cross section for fragmentation, as compared to the case of an identified jet. 

The factorization formula for the inclusive production of an identified jet can be written as
\begin{align}
\label{eq:HJ}
\frac{d\sigma}{dz_J} =&\int dx\,dz~ \vec{H}\left(x,\frac{Q^2}{\mu^2},\mu \right) \cdot \vec{J}\left(z,\ln\frac{x^2 Q^2R^2}{4\mu^2},\mu \right) \delta(z_J - x z)\\
=&\int_{z_J}^1 \frac{dx}{x}~ \vec{H}\left(x,\frac{Q^2}{\mu^2},\mu \right) \cdot \vec{J}\left(\frac{z_J}{x},\ln\frac{x^2 Q^2 R^2}{4\mu^2},\mu \right)\,,\nn
\end{align}
where $z_J = 2E_J/Q$ is the energy fraction of the observed jet. 
Here $\vec{H}$ is again the standard inclusive hard function for hadron fragmentation in \eq{frag}, which obeys the timelike DGLAP equation, and the (``semi-'')inclusive jet function $\vec{J}$ is defined for quark and gluon jets in terms of gauge-invariant quark and gluon fields in SCET as
\begin{align}
J_q\left(z=p_J^- / \omega, \omega_J, \mu\right)&=\frac{z}{2 N_c} \operatorname{Tr}\left[\frac{\slashed{\bar{n}}}{2}\left\langle 0\left|\delta(\omega-\bar{n} \cdot \mathcal{P}) \chi_n(0)\right| J X\right\rangle\left\langle J X\left|\bar{\chi}_n(0)\right| 0\right\rangle\right]\,, \\
J_g\left(z=p_J^- / \omega, \omega_J, \mu\right)&=-\frac{z \omega}{2\left(N_c^2-1\right)}\left\langle 0\left|\delta(\omega-\bar{n} \cdot \mathcal{P}) \mathcal{B}_{n \perp \mu}(0)\right| J X\right\rangle\left\langle J X\left|\mathcal{B}_{n \perp}^\mu(0)\right| 0\right\rangle\,.\nn
\end{align}
Here $J$ and $X$ denote the identified jet, and the unidentified state, respectively. The operators $\chi$ and $\mathcal{B}$ are the standard gauge-invariant collinear quark and gluon fields in SCET \cite{Bauer:2001yt,Bauer:2002nz,Bauer:2001ct,Bauer:2000yr}, and the label operator $\bar n \cdot \mathcal{P}$ measures the large component of the momentum.

We see that the case of identified jet production has a new convolution structure relative to the hadron fragmentation case that is reminiscent of that of energy correlators, where extra convolution variables $x$ dependence show up in the logs. Indeed, this should be intuitively clear, since we are defining jets with an angular scale $R^2$, similar to the energy correlators with $x_L$. 

The new convolution structure modifies the RG evolution equation for the inclusive jet function relative to that of the hadron fragmentation function in \eq{hadRG} as
\begin{align}
  \label{eq:jetRG}
 \frac{d \vec{J}\left(z,\ln\frac{Q^2 R^2}{4\mu^2},\mu \right)}{d \ln \mu^2} = \int_z^1 \frac{dy}{y}  \vec{J}\left(\frac{z}{y},\ln\frac{y^2 Q^2 R^2}{4\mu^2},\mu \right) \cdot \widehat P_T(y)\,,
\end{align}
which is not the standard DGLAP form. Taking the $N$-th moments of \eq{HJ}, we find that the convolution structure is exactly like that of the projected $N$-point projected energy correlator,\footnote{Here, $N$ does not have to be an integer. The analytic continuation of $N$-point projected energy correlator was considered in~\cite{Chen:2020vvp} and will also be discussed in sec.~\ref{sec:renorm}.}
\begin{align}
\int_0^1 dz_J\,z_J^{N}\,\frac{d\sigma_h}{dz_J} =& \int_0^1 dx\,x^{N}\vec{H}\left(x,\frac{Q^2}{\mu^2},\mu \right) \cdot \vec{J}\left(N,\ln\frac{x^2 Q^2 R^2}{4\mu^2},\mu \right)\,,
\end{align}
where the $N$-th moment of the inclusive jet function is defined as
\begin{align}
\label{eq:hadDGmom}
\vec{J} (N,\ln\frac{Q^2 R^2}{4\mu^2},\mu) &\equiv \int_0^1 dz\, z^{N}\,\vec{J} (z,\ln\frac{Q^2 R^2}{4\mu^2},\mu)\,.
\end{align}
The RG consistency of the factorization formula and the modified convolution structure relative to the hadron fragmentation process then implies that $N$-th moment of the inclusive jet function also follows the RG structure of the $N$-point projected energy correlators as 
\begin{align}
  \label{eq:incljetRG}
\frac{d \vec{J}(N,\ln\frac{Q^2R^2}{4\mu^2}, \mu) }{d \ln \mu^2} = \int_0^1 dy\, y^{N} \vec{J} (N,\ln\frac{y^2 Q^2R^2}{4\mu^2}, \mu) \cdot \widehat P_T(y,\mu) \,.
\end{align}
As emphasized for the energy correlators, this is not DGLAP, but is completely determined by the standard timelike splitting functions. The identical factorization structure between the moments of inclusive jets and the $N$-point projected correlators~\cite{Chen:2020vvp,Dixon:2019uzg} suggests that inclusive jets can be viewed as a generalization of the factorization theorems for energy correlators, encompassing all moments.

It is important to emphasize that this factorization theorem is at the same level of rigor as the Collins-Soper-Sterman factorization theorem for inclusive hadron production \cite{Collins:1981ta,Bodwin:1984hc,Collins:1985ue,Collins:1988ig,Collins:1989gx,Nayak:2005rt,Collins:2011zzd}, since it is only a modification of the IR dynamics of the jet function. This is important, particularly to its application to precision calculations in proton-proton collisions.

To understand how this factorization theorem works, we can now perform the same exercise as we did above for the energy correlator, to see that when we identify a jet, their moments also exhibit power-law scaling of $R^2$ by the spacelike anomalous dimension by the reciprocity and modified convolution structure. 

Therefore, quite interestingly, we find that identifying a jet using the anti-$k_T$ (or generalized $k_T$ family) algorithm, flips the anomalous dimension from timelike to spacelike, as compared to identified hadron production. We find it elegant, that this story, which was understood in the context of the energy correlators, using advances from conformal field theory, has applications in the much more phenomenological context of jet production. We believe that this is an important advance in the understanding of timelike factorization.

The general convolution structure of our factorization theorem trivially generalizes to a number of other cases of interest for collider physics, and we will discuss these in more detail in \Sec{sec:fact_gen}. 

\subsection{Comparison With Other Approaches}\label{sec:fact_comment}
In this section, we briefly comment on the relation between our new factorization theorem, and other approaches in the literature.

We first compare with the formulations from \cite{Kang:2016mcy}. Since this approach is also formulated as a factorization formula in terms of definite matrix elements in SCET, this comparison is most straightforward. In \cite{Kang:2016mcy}, the following factorization formula was proposed for jet energy distribution in $e^+e^-$ as
\begin{align}
\label{eq:old}
\frac{d\sigma}{dz_J}\bigg|_{\text{\cite{Kang:2016mcy}}} =&\int dx\,dz~ \vec{H}\left(x,\frac{Q^2}{\mu^2},\mu \right) \cdot \vec{J}\left(z,\ln\frac{Q^2 R^2}{4\mu^2},\mu \right) \delta(z_J - x z)\\
=&\int_{z_J}^1 \frac{dx}{x}~ \vec{H}\left(x,\frac{Q^2}{\mu^2},\mu \right) \cdot \vec{J}\left(\frac{z_J}{x},\ln\frac{Q^2 R^2}{4\mu^2},\mu \right)\,,\nn
\end{align}
and the corresponding formulation for high-$p_T$ jet as discussed around \eq{highpTfrag}. Renormalization group consistency implies that the fragmenting jet function obeys \emph{exactly} the timelike DGLAP equation. Our explicit two-loop calculation shows that this is incorrect, and that the above factorization formula is therefore invalid. The fact that this factorization formula is not correct was also found in \cite{vanBeekveld:2024jnx} by explicit calculation.

At LL accuracy, this factorization is equivalent to that presented in this paper. This is no longer the case already at NLL. This is similar to the case of the factorization theorem for the energy correlator presented in \cite{Dixon:2019uzg}, where older approaches in terms of the jet calculus \cite{Konishi:1979cb} did not incorporate correct convolutions with hard functions.

The reason the simplified convolution structure of \cite{Kang:2016mcy,Kang:2016ehg} is insufficient to describe a measurement defined by a jet algorithm can be understood as follows. In all the factorization theorems above, at leading power we factorize onto massless single particle states, with well-defined quantum numbers under the Lorentz group, in particular boosts. In the case of single identified hadron production, these quantum numbers match those of convolution variable in the fragmentation function, allowing them to be directly tied together by a convolution. However, the energy fraction carried by an identified jet with a jet algorithm, and jet radius, has different properties. This variable therefore cannot be directly tied together as a convolution with that of the intermediate single particle state, as in Eq.~\eqref{eq:old}, since it has inconsistent transformation properties under boosts. In this case an additional variable is required, leading to convolution structure of the factorization theorem presented in this paper.

Similar conclusions were also presented in \cite{vanBeekveld:2024jnx}, and further explored in \cite{vanBeekveld:2024qxs}, and we agree with the calculations presented there, as well as their conclusion that the logarithmic structure of the inclusive jet function is modified. However, we would like to highlight several differences in our interpretation of this modified logarithmic structure, as well as highlight several features of the all-orders organization achieved by the factorization theorem presented here, which we find to be advantageous. First, we wish to emphasize that the modification of the jet algorithm, being a modification of the IR measurement, does not modify the anomalous dimension, which is a UV quantity. Although one can define an effective anomalous dimension order by order in perturbation theory, as in \cite{vanBeekveld:2024jnx}, this does not represent the true anomalous dimension of any object in quantum field theory. We clearly illustrated this using the example of the energy correlator in the discussion surrounding \Eq{eq:pureYMNLL}, where we showed how to get the same result from an effective anomalous dimension, and a modified RG equation. While this may seem like a minor technicality in interpretation, the key advantage of our formulation is that it demonstrates, to all orders in perturbation theory, that the hard function for inclusive jets is the same universal hard function as for hadron fragmentation, and that the RG evolution of the jet function is entirely determined by DGLAP, although in a non-trivial manner. In other words, the effective anomalous dimension is not an independent parameter of the field theory, but rather is fixed by DGLAP. This is important, as it allows us to extend our results to higher orders using known DGLAP anomalous dimensions.

It is also important to clarify that our observations about a modified RG structure only apply to the case of inclusive jets, and we are not claiming such a modification for exclusive factorization theorems. The case of exclusive jets exhibits a structurally different Sudakov-style factorization~\cite{Ellis:2010rwa,Chien:2015cka,Bertolini:2017efs}. However, as a byproduct of our calculation, we will present new results for the gluon exclusive jet function in sec.~\ref{sec:excljet}.

\section{Generalizations to Jet Substructure Factorization}\label{sec:fact_gen}
The inclusive jet function incorporates all aspects of the clustering measurement at the jet scale $Q\,R/2$, fully factorizing it from the dynamics at the hard scale $Q$. Since this factorization between the hard and jet scales remains intact even when jet substructures are measured, we only need to modify the jet functions for different jet substructure observables. Specifically, for arbitrary differential jet substructure measurements $\{v_1,v_2,\cdots,v_n\}$, the inclusive jet function can be replaced by the substructure-sensitive jet function as follows
\begin{align}
\label{eq:jetrepl}
J_i(z,R,\mu) \to \mathcal{G}_i (z,R,v_1,v_2,\cdots,v_n,\mu)\,.
\end{align}
This simple structure has facilitated numerous extensions of the inclusive jet framework to a wide range of jet substructure observables~\cite{Hannesdottir:2022rsl,Lee:2023xzv,Craft:2022kdo,Chien:2018lmv,Lee:2022ige,Lee:2023npz,Aschenauer:2019uex,Kang:2018qra,Cal:2020flh,Kang:2020xyq,Kang:2018vgn,Kang:2018jwa,Kang:2019prh,Cal:2021fla,Cal:2019gxa,Kang:2016ehg,Cal:2019hjc,Kang:2017mda,Mehtar-Tani:2024smp}. 

In this section, we extend the discussion of the modified convolution structure in the factorization formula for inclusive jet production to its implications for jet substructure observables. We can broadly categorize different jet substructures into two types: collinear observables, which involve only hard-collinear factorization, and soft-sensitive observables, which involve soft-collinear factorization. We will discuss the impact of our factorization approach in both categories and examine specific cases within each to show how the structure of $\mathcal{G}_i$ in~\eq{jetrepl} is modified. For simplicity and convenience in comparison with the discussion above, we will continue to discuss factorization theorems using the case of $e^+e^-$ with jet energy fraction. Generalization to high-$p_T$ jet production case follows immediately as discussed around~\eq{highpTfrag}.

\subsection{Collinear Jet Substructure Observables}\label{sec:fact_coll}

We begin by discussing collinear jet substructure observables, which are sensitive only to collinear dynamics. The prototypical example is the measurement of the energy fraction of an identified hadron within a jet, which was the first jet substructure extension of inclusive jet production factorization. In~\cite{Kang:2016ehg}, a hard-collinear factorization formula building on the semi-inclusive jet function approach \cite{Kang:2016mcy} was developed to account for the energy fraction of the hadron relative to the jet energy in jets produced through single inclusive production. Since our factorization theorem modifies the convolution structure in the semi-inclusive jet function, it also modifies the convolution structure in this broader class of factorization theorems. Here we will show how our newly introduced factorization theorem extends to this case. There are many extensions and generalizations that use the factorization theorem for identified hadron production within a jet as a key building block. These include
\begin{itemize}
\item Identified (multi-)hadron production (fragmenting jet functions) \cite{Kang:2016ehg,Lee:2023xzv}
\item Polarized fragmenting jet functions \cite{Kang:2020xyq,Kang:2023elg}
\item Energy correlators in jets \cite{Lee:2022ige}
\item Track energy fraction in jets \cite{Lee:2023xzv}
\end{itemize}
This shows the breadth of applications of our newly developed factorization theorem. Here we only highlight the structure of the factorization theorem, but it would be interesting to investigate the phenomenological implications of our modified factorization in each of these cases.

The factorization formula for the inclusive production with identified hadron energy fraction can be written as
\begin{align}
\label{eq:hadinjet}
\frac{d\sigma^h}{dz_J dz_h} =&\int_{z_J}^1 \frac{dx}{x}~ H_i\left(x,\frac{Q^2}{\mu^2},\mu \right)  \mathcal{G}_{i\to h}\left(\frac{z_J}{x},z_h,\ln\frac{x^2 Q^2 R^2}{4\mu^2},\mu \right)\,,\nn\\
=&\int_{z_J}^1 \frac{dx}{x}~ H_i\left(x,\frac{Q^2}{\mu^2},\mu \right)  \left[\int_{z_h}^1 \frac{dy}{y}\mathcal{J}_{ij}\left(\frac{z_J}{x},y,\ln\frac{x^2 Q^2 R^2}{4\mu^2},\mu \right) D_{j\to h}\left(\frac{z_h}{y},\mu\right)\right]\,,
\end{align}
where $z_h = E_h/E_J$ is the energy fraction with respect to the jet energy. The first equation illustrates that the jet substructure sensitive, in this case hadron energy fraction in jet, case begins simply by replacing the inclusive jet function with the corresponding substructure sensitive jet function as alluded to in~\eq{jetrepl}. This new hadron momentum fraction sensitive jet function $\mathcal{G}_{i\to h}$ can then be matched to the universal collinear fragmentation function $D_{j\to h}$ as illustrated by the factor in the bracket $[\cdots]$ in the second line. Here, $\mathcal{J}_{ij}$ is the collinear matching coefficient sensitive to both energy fraction of the jet with respect to initiating parton $i$ and energy fraction of the parton $j$ with respect to the jet, given by the first and second arguments, respectively. Relative to the factorization given in~\cite{Kang:2016ehg}, we see modified convolution structure in $\mathcal{J}_{ij}$ in the $x$ convolution. This is contrary to the finding in~\cite{Kang:2016ehg} that $x$ has the usual timelike DGLAP convolution structure. This is therefore a fun example of a factorization, where due to reciprocity, it exhibits both spacelike scaling in $z$ convolution for jet production and usual timelike DGLAP in $x$ convolution for hadron production. Extension to the case involving polarized hadron in jet~\cite{Kang:2020xyq} is straightforward.

The identified single hadron-in-jet factorization can be generalized to the case involving multiple energy fractions for the intermediate states that initiating parton $i$ splits to. An example of such observable sensitive to all of the energy fractions of individual intermediate state is track energy fraction inside jets. As tracks can come from all the intermediate branches of splitting, maintaining a convolution structure from all the splitting is important. The relevant factorization framework was worked out recently in~\cite{Lee:2023xzv}. This factorization is now modified as
\begin{align}
\frac{d\sigma^{\rm trk}}{dz_J dz_{\rm trk}} =&\int_{z_J}^1 \frac{dx}{x}~ H_i\left(x,\frac{Q^2}{\mu^2},\mu \right)  \mathcal{G}_{i\to \rm trk}\left(\frac{z_J}{x},z_{\rm trk},\ln\frac{x^2 Q^2 R^2}{4\mu^2},\mu \right)\,,\nn\\
=&\int_{z_J}^1 \frac{dx}{x}~ H_i\left(x,\frac{Q^2}{\mu^2},\mu \right)  \bigg[\sum_{m=1} \int \prod_{k=1}^m\left[dy_k dx_k T_{i_k}(x_k)\right]
\delta\left(z_{\rm{trk }}-\sum_{k=1}^m x_k y_k\right) \nn\\
&\hspace{1cm}\times\delta\left(\sum_{k=1}^m y_k-1\right)\mathcal{J}_{i \rightarrow\left[i_1, \ldots, i_m\right]}\left(\frac{z_J}{x}, y_1, \ldots, y_{m-1}, \ln\frac{x^2 Q^2 R^2}{4\mu^2}, \mu\right)\bigg]\,,
\end{align}
where  $z_{\rm trk} = E_{\rm trk}/E_J$ is the energy fraction of the track with respect to the jet energy and $\mathcal{J}_{i\to [i_1,\cdots,i_m]}$ is the collinear matching coefficients sensitive to the energy fraction of all $m$ intermediate partons with respect to the jet, as well as the energy fraction of the jet with respect to the initiating parton $i$. These $m$ partons then turn into tracks, which is described by the product of $m$ track functions. The non-perturbative track function $T_j(x)$ gives the probability density of parton $j$ fragmenting into tracks with energy fraction $x$~\cite{Chen:2022muj,Jaarsma:2022kdd,Li:2021zcf,Chang:2013iba,Chang:2013rca,Chen:2022pdu,Lee:2023tkr}. Relative to the factorization presented in~\cite{Lee:2023xzv}, we again see the matching coefficients have modified convolution structure in the $x$ convolution. This factorization then gives non-linear track function evolution~\cite{Chen:2022muj,Chen:2022pdu,Jaarsma:2022kdd} from where the track function is extracted to the jet scale $Q\,R/2$, which then gives inclusive jet RG between the jet scale and the hard scale. As described in Sec 2.4 of~\cite{Lee:2023xzv}, the track function case can straightforwardly be generalized to the factorization involving multi-hadron fragmentations in jet. The convolution variable describing the jet production is similarly modified as here for such multi-hadron case.

Another very interesting case involving collinear dynamics is observing energy correlator within a jet. This follows a similar hard-collinear factorization described around~\eq{fact_N} derived in~\cite{Dixon:2019uzg}, except now we are using jet energy as the reference scale to calculate energy correlators. Having a reference jet scale is necessary in the case of hadron colliders, where we cannot tag the initial scale of the collision. This factorization with respect to the jet scale was stated for the first time in~\cite{Lee:2022ige}, which should be corrected due to the modified convolution structure for jet production variable.  Factorization is simple to state for the cumulant, which we define as
\begin{align}
\Sigma^{[N]}\left(x_L,z_J\right)\equiv\frac{1}{\sigma} \int_0^{x_L} d x_L^{\prime} \frac{d \sigma^{[N]}}{d x_L^{\prime}dz_J}
\end{align}
for projected $N$-point energy correlators. Then in the small angle limit, the cumulant exhibits factorization as
\begin{align}
\Sigma^{[N]}(x_L,z_J)=&\int_{z_J}^1 \frac{dx}{x}~ H_i\left(x,\frac{Q^2}{\mu^2},\mu \right)  \mathcal{G}_{i}\left(\frac{z_J}{x},x_L,\ln\frac{x^2 Q^2 R^2}{4\mu^2},\mu \right)\,,\\
=&\int_{z_J}^1 \frac{dx}{x}~ H_i\left(x,\frac{Q^2}{\mu^2},\mu \right)  \left[\int_{0}^1 dy\, y^N\,\mathcal{J}_{ij}\left(\frac{z_J}{x},y,\ln\frac{x^2 Q^2 R^2}{4\mu^2},\mu \right) J^{[N]}_{j}(\ln\frac{x_Ly^2 Q^2}{\mu^2},\mu)\right]\,.\nn
\end{align}
This case is very interesting as both the inclusive jet production and energy correlator as jet substructure exhibit spacelike scaling due to modified convolution structure with respect to the usual DGLAP convolution. To make this more transparent, we can take the moments in $z_J$ and derive
\begin{align}
\int dz_J\,z_J^M\,\Sigma^{[N]}(x_L,z_J)=&\int_{0}^1 dx\, x^M\, H_i\left(x,\frac{Q^2}{\mu^2},\mu \right) \\ &\times\left[\int_{0}^1 dy\, y^N\,\mathcal{J}_{ij}\left(M,y,\ln\frac{x^2 Q^2 R^2}{4\mu^2},\mu \right) J^{[N]}_{j}(\ln \frac{x_Ly^2 Q^2}{\mu^2},\mu)\right]\,,\nn
\end{align}
where
\begin{align}
\mathcal{J}_{ij}\left(M,y,\ln\frac{x^2 Q^2 R^2}{4\mu^2},\mu \right)\equiv \int_0^1 dz z^M \mathcal{J}_{ij}\left(z,y,\ln\frac{x^2 Q^2 R^2}{4\mu^2},\mu \right)\,.
\end{align}
This makes the mirror structure between the inclusive jet function moments and projected $N$-point energy correlator more apparent. In particular, we can repeat the exercise by using reciprocity to demonstrate that the jet function evolves by spacelike anomalous dimension with spin-$N$ from the energy correlator scale $\mu_{\rm EEC}\sim Q\sqrt{x_L}$ to the jet scale $\mu_{J}\sim Q R/2$, which then evolve by spacelike anomalous dimension of spin-$M$ to the hard scale $\mu_{H}\sim Q$.

It is fun to phrase this again as a modification of the IR measurement, as compared to the hadron energy fraction in jet case discussed in~\eq{hadinjet}. In the hadron-in-jet case, we had spacelike evolution in the jet fraction $z_J$ and timelike evolution in the hadron fraction $z_h$. Now, we modify the IR measurement, but at a scale much below $Q\,R/2$. Therefore the spacelike evolution in $z_J$ is not modified, but we get the exact same reciprocity for IR measurement, which now flips it to a spacelike scaling in the angle of the energy correlator!

This can also be extended to the case of energy correlators measured on a massive quark, which is important phenomenologically for both the bottom and top quarks
\cite{Craft:2022kdo,Holguin:2023bjf,Holguin:2022epo}. In this case, the quark mass, being an IR effect, does not appear in the hard function, and therefore does not modify the convolution structure presented in this paper.

From the perspective of phenomenological applications, there are a number of nice features of this equation, that were highlighted in \cite{Lee:2022ige}. First, the impact of the jet clustering occurs only at the perturbative scale $Q\,R/2$ and therefore does not interfere with either the perturbative, or non-perturbative structure \cite{Chen:2024nyc,Lee:2024esz} of the energy correlator measurement at lower scale. The factorization theorem presented here will be key to extending the precision jet substructure program at the LHC.

Finally, another interesting case involving only collinear dynamics is the case of leading and subleading jets considered in~\cite{Scott:2019wlk,Neill:2021std}. The factorization there involves multiple momentum fractions from the hard function as it involves multiple jet production. This can be thought of as extending our formalism to the case of double, triple, $\cdots$, -inclusive jet production case and imposing phase space cuts to study the leading or subleading jet spectrum. We will not explicitly consider the factorization of this kind as the factorization was only considered to leading logarithms in $\ln R$ in the literature, but the convolution structure for production of jets will similarly modify the convolution structure with respect to the hadron case.

\subsection{Soft-sensitive Jet Substructure Observables}\label{sec:fact_coll}
Within the single inclusive jet production framework, there have been many soft-sensitive jet substructure studies~\cite{Hannesdottir:2022rsl,Chien:2018lmv,Aschenauer:2019uex,Kang:2018qra,Cal:2020flh,Kang:2018vgn,Kang:2018jwa,Kang:2019prh,Cal:2021fla,Cal:2019gxa,Cal:2019hjc}. In general, the first step of the factorization still follows the hard-collinear factorization where we can simply replace the inclusive jet function by the substructure-sensitive jet function as in~\eq{jetrepl}. Then for general substructure observable, we have a convolution structure
\begin{align}
\label{eq:vsub}
\frac{d\sigma}{dz_J dv} =&\int_{z_J}^1 \frac{dx}{x}~ H_i\left(x,\frac{Q^2}{\mu^2},\mu \right)  \mathcal{G}_{i}\left(\frac{z_J}{x},v,\ln\frac{x^2 Q^2 R^2}{4\mu^2},\mu \right)\,,
\end{align}
where $v$ is some soft-scale sensitive substructure we measure. This hard-collinear factorization is valid for fixed-order region of the $\mathcal{G}_{i}$ sensitive to $v$. In general, such jet substructure measurements then give sensitivity to some scale sensitive to $v$ measurement, $\mu_v$. In the region where $\mu_v \ll Q\,R/2$, we can further carry out a factorization
\begin{align}
\label{eq:Gmatching}
\mathcal{G}_{i}\left(\frac{z_J}{x},v,\ln\frac{x^2 Q^2 R^2}{4\mu^2},\mu \right) = \mathcal{H}_{i\to j}\left(\frac{z_J}{x},\ln \frac{x^2Q^2R^2}{4\mu^2},\mu\right)\mathcal{V}_j(v,\mu)\left(1+\mathcal{O}(2\mu_v/ Q R)\right)\,.
\end{align}
Here, $\mathcal{V}$ denotes the jet function sensitive to measurement of the jet substructure $v$. Typically, there is sensitivity to both soft and collinear scales due to such jet substructure measurements, and thus $\mathcal{V}$ can be factorized in terms of soft and collinear functions describing their respective dynamics. On the other hand, $\mathcal{H}_{i\to j}$ is the hard matching function at the jet scale $Q\,R/2$.  As discussed in~\cite{Kang:2019prh}, note that jet substructure observables can in general introduce NGL contributions between this hard matching function and $\mathcal{V}_j$. Relative to all the matching procedures described in the literature, the convolution structure of the hard matching coefficients must be modified as~\eq{Gmatching}.

\section{A Jet Algorithm Definition of the $\nu$-Point Generating Function}\label{sec:EECalg}

In sec.~\ref{sec:fact}, we studied the intimate connection between the factorization structure and the RG structure of the moments of inclusive jet functions and projected energy correlator jet functions. In particular, we found that the $N$-th moment of the inclusive jet function has exactly the same factorization and RG structure as the $N$ point projected energy correlator jet functions. Therefore, up to constants unconstrained by RG, we expect the two jet functions to be exactly identical. Remarkably, this then implies that the inclusive jet productions of any jet algorithm that exhibit the factorization structure we described act almost as a \emph{generating functional} for projected energy correlators, where the distribution in $z$ encodes the complete information of the projected energy correlators for any $N$. The $N$-th moment of the inclusive jet function then projects to the desired $N$-point projected energy correlator, again up to constants.

In fact, at NLO, even the constants associated with $k_T$-type jet algorithms match precisely with the one-loop jet functions for the projected energy correlators. This indicates that inclusive jets for $k_T$-type jet algorithms serve \emph{exactly} as the generating functional for projected energy correlators at NLL accuracy. However, this correspondence between $k_T$-type jet algorithms and projected energy correlators breaks down at two-loop orders due to differences in the two-loop constants. This motivates us to propose a completely new jet algorithm whose jet function serves as the exact generating functional for projected energy correlators to arbitrary perturbative order, which is the goal of this section.

We begin by reviewing the measurements of the energy correlator jet function. In~\cite{Chen:2020vvp}, the analytic continuation of the $N$-point projected energy correlator, referred to as the $\nu$-correlators for short, was introduced, where $\nu$ can take non-integer values. It was also demonstrated that these $\nu$-correlators are IRC safe. At fixed order, the cumulants for the $\nu$-correlators are given as
\begin{align}
\label{eq:projected_mom_nu}
\Sigma^{[\nu]} (x_L)&\ = \sum_n \int \! d\sigma_{X_n} 
\cdot 
\Big[
\sum_{1\leq i_1 \leq n} \cW_1^{[\nu]}(i_1) \Theta(x_L)  +
\sum_{1 \leq i_1 < i_2 \leq n} \cW_2^{[\nu]} (i_1, i_2) \Theta(x_L - z_{i_1 i_2} )
\nn\\
&\ + 
\sum_{1 \leq i_1 < i_2 < i_3 \leq n} \cW_3^{[\nu]} (i_1, i_2, i_3) \Theta(x_L - \max \{z_{i_1 i_2} , z_{i_1 i_3}, z_{i_2 i_3} \} ) + \ldots
\nn\\
&\ + 
\sum_{1 = i_1 < i_2 < \ldots < i_n = n} \cW_n^{[\nu]} (i_1, i_2, \ldots, i_n) \Theta(x_L - \max \{z_{i_1 i_2} , z_{i_1 i_3}, \ldots, z_{i_{n-1} i_n}  \} ) \Big]  \,.
\end{align}
Although the definition is provided for the cumulants of $\nu$-correlators, the $\nu$-correlator jet function for small angles $x_L$ is defined using similar measurement functions $\cW$. For a fixed number of final states $n$, each term $\cW_m^{[\nu]}$ considers correlations among a subset of particles, where $m \leq n$. Theta functions are applied to ensure that the largest angle remains smaller than $x_L$ for the cumulants. Finally, the weights $\cW_m^{[\nu]}$ corresponding to different particle correlations are given as
\begin{align}
  \label{eq:weight_func}
 \cW_1^{[\nu]} (i_1) = &\ x_{i_1}^\nu \,
\nn\\
  \cW_2^{[\nu]} (i_1, i_2) = &\ \left(x_{i_1}+x_{i_2}\right)^\nu - \sum_{1 \leq a \leq 2} \cW_1^{[\nu]}(i_a) \,,
\nn\\
\cW_3^{[\nu]}(i_1, i_2, i_3) = &\ \left(\sum_{a=1}^3 x_{i_a}\right)^\nu - \sum_{1 \leq a < b \leq 3} \cW_2^{[\nu]}(i_a, i_b) - \sum_{1 \leq a \leq 3} \cW_1^{[\nu]}(i_a) \,,
\\
\ldots &\  \,,
\nn\\
  \cW_n^{[\nu]} (i_1, \ldots, i_n) = &\ \left(\sum_{a=1}^n x_{i_a}\right)^\nu 
- \sum_{1 \leq a_1 < a_2 < \ldots < a_{n-1} \leq n} \cW_{n-1}^{[\nu]}(i_{a_1} , i_{a_2} , \ldots , i_{a_{n-1}})
- \ldots - \sum_{1 \leq a \leq n }^n \cW_1^{[\nu]}(i_a) \,.\nn
\end{align}
Now, we want to define a jet algorithm that maps exactly to such measurements when moments are taken. This energy correlator inspired jet algorithm at fixed order can simply be defined as  
\begin{align}
  \label{eq:EECalg}
\frac{d\sigma}{dz_J}&\ = \sum_n \int \! d\sigma_{X_n} 
 \cdot 
\Big[
 \sum_{1\leq i_1 \leq n} \cV_1(i_1) \Theta(x_L)  +
\sum_{1 \leq i_1 < i_2 \leq n} \cV_2 (i_1, i_2) \Theta(x_L - z_{i_1 i_2} )
\nn\\
&\ + 
\sum_{1 \leq i_1 < i_2 < i_3 \leq n} \cV_3 (i_1, i_2, i_3) \Theta(x_L - \max \{z_{i_1 i_2} , z_{i_1 i_3}, z_{i_2 i_3} \} ) + \ldots
\nn\\
&\ + 
\sum_{1 = i_1 < i_2 < \ldots < i_n = n} \cV_n (i_1, i_2, \ldots, i_n) \Theta(x_L - \max \{z_{i_1 i_2} , z_{i_1 i_3}, \ldots, z_{i_{n-1} i_n}  \} ) \Big]  \,,
\end{align}
where
\begin{align}
  \label{eq:jetweight_func}
  \cV_1 (i_1) = &\ \delta(z_J - x_{i_1}) \,
\nn\\
  \cV_2 (i_1, i_2) = &\ \delta(z_J-\left(x_{i_1}+x_{i_2}\right)) - \sum_{1 \leq a \leq 2} \cV_1(i_a) \,,
\nn\\
\cV_3(i_1, i_2, i_3) = &\ \delta(z_J - \sum_{a=1}^3 x_{i_a}) - \sum_{1 \leq a < b \leq 3} \cV_2(i_a, i_b) - \sum_{1 \leq a \leq 3} \cV_1(i_a) \,,
\\
\ldots &\  \,,
\nn\\
  \cV_n (i_1, \ldots, i_n) = &\ \delta(z_J-\sum_{a=1}^n x_{i_a}) 
- \sum_{1 \leq a_1 < a_2 < \ldots < a_{n-1} \leq n} \cV_{n-1}(i_{a_1} , i_{a_2} , \ldots , i_{a_{n-1}})
- \ldots - \sum_{1 \leq a \leq n }^n \cV_1(i_a) \,.\nn
\end{align}
It is straightforward to see that taking the $\nu$ moment of this inclusive jet cross section with new algorithm directly maps to the $\nu$-correlators, and thus encapsulates information about all $\nu$ correlators for any arbitrary $\nu$ simultaneously. While experimental procedure to measure this jet cross section is obvious from definition, a few points are worth noting for the physical interpretation of this jet algorithm. First, the fixed value of $x_L$ functions as a jet radius-like parameter. This is obvious as the theta functions restrict the maximum angle to be no larger than $x_L$, effectively considering a set of particles clustered within a region of approximate size $x_L$ (or more precisely, within the angular size $\theta\sim 2\sqrt{x_L}$ as $x_L$ is being compared to $z_{ij} = (1-\cos\theta_{ij})/2 \approx \theta_{ij}^2/4$). We then examine all possible combinations of these particles into different subsets to measure energy fractions. These can be thought of as considering jets formed by smaller subsets of final state particles. As we will see below for $k_T$-type algorithm, this is a common feature for inclusive jet measurements. Due to its connection to $\nu$-correlators, our new jet algorithm is also immediately IRC safe.

It is also interesting for us to compare our new jet algorithm to $k_T$ type jet algorithm for one-loop jet functions. In \Sec{sec:calc}, we will present the complete measurement function for $k_T$-type jet algorithms at two-loop. For the one-loop jet function, there can be up to two particles in the final state, leading to only two possible topologies: either both particles are clustered into a single jet, or they are separately clustered into different jets. The relevant jet measurements can then be represented as~\cite{Kang:2016mcy}
\begin{align}
\label{eq:twoparticlemeas}
&\delta(z_J - (x_1 + x_2)) \theta\left(\frac{R^2}{4}-z_{12}\right) + \left[\delta(z_J - x_1)+\delta(z_J - x_2)\right] \theta\left(z_{12}-\frac{R^2}{4}\right)\nn\\
&=\bigg[\delta(z_J - (x_1 + x_2)) - \delta(z_J - x_1)-\delta(z_J - x_2)\bigg] \theta\left(\frac{R^2}{4}-z_{12}\right) +\delta(z_J - x_1)+\delta(z_J - x_2) \nn\\
&= \mathcal{V}_2(1,2) \theta\left(\frac{R^2}{4}-z_{12}\right) + \left(\mathcal{V}_1(1) + \mathcal{V}_1(2)\right) \theta\left(\frac{R^2}{4}\right)\,,
\end{align}
which exactly matches our new inclusive jet algorithm for two-particle final state $n=2$ case. That is, our new jet algorithm and $k_T$-type jet algorithms are identical to one-loop accuracy at fixed order, as well as to the NLL resummation accuracy. To this accuracy, therefore, $k_T$-type jet function is exactly the generating functional for the $\nu$-correlators, simultaneously encapsulating all $\nu$-correlators information with $R^2/4$ identified with $x_L$. The NLO jet function was explicitly computed as~\cite{Kang:2016mcy,Neill:2021std}
\begin{align}\label{eq:inclusiveJfunction}
 J_q(z,E R,\mu) = 
 & \, 
 \delta(1-z)+\f{\as}{2\pi}\left(\ln\left(\f{\mu^2}{E^2 R^2}\right)-2\ln z\right)\left[P_{qq}(z)+P_{gq}(z)\right]
 \nn \\&
 -\f{\as}{2\pi}\Bigg[C_F\left[2(1+z^2)\left(\f{\ln(1-z)}{1-z}\right)_++(1-z)\right]
 \nn\\&
-\delta(1-z) C_F\left(\f{13}{2}-\f{2\pi^2}{3}\right)+2 P_{gq}(z) \ln(1-z)+C_F z\Bigg]
\\
J_g(z,E R,\mu) = 
& \, 
\delta(1-z) + \f{\as}{2\pi}\left(\ln\left(\f{\mu^2}{E^2 R^2}\right)-2\ln z\right)\left[P_{gg}(z)+2 N_f P_{qg}(z)\right]
\nn \\&
-\frac{\alpha_{s}}{2 \pi}\Bigg[\frac{4 C_{A}\left(1-z+z^{2}\right)^{2}}{z}\left(\frac{\ln (1-z)}{1-z}\right)_{+}-\delta(1-z)\bigg(C_{A}\left(\frac{67}{9}-\frac{2 \pi^{2}}{3}\right)
\nn\\&
-T_{F} N_{f}\left(\frac{23}{9}\right)\bigg)+4 N_{f}\left(P_{q g}(z) \ln (1-z)+T_{F} z(1-z)\right)\Bigg]\,,
\end{align}
where $E$ denotes the energy or the transverse momentum of the jet initiating parton. Note that $E$ will also be denoted as $Q/2$ below at the jet function level. The $\nu$ moment matches exactly the $\nu$ correlator expression presented in~\cite{Chen:2020vvp}. As we will see, this correspondence is broken explicitly at two-loop. It would be very interesting to study the inclusive jet function of the new jet algorithm defined in \eq{EECalg} to higher precision, which would simultaneously contain information for arbitrary $\nu$-point energy correlators at any perturbative accuracy. 

\section{Renormalization Group Structure}\label{sec:renorm}
We now use the renormalization group analysis to predict the structure of the bare semi-inclusive jet function to two loops. We will then compare this prediction with the numerical calculations in the next section. For convenience, we define the shorthand notations for the convolutions involved as
\begin{align}
[A\otimes B](z) &\equiv \int_z^1 \frac{dy}{y} A(y) B(z/y) = \int_z^1 \frac{dy}{y} A(z/y) B(y)\,,\nn\\
[A \boxtimes B]\left(z,\ln \frac{Q^2R^2}{4\mu^2}\right) &\equiv \int_z^1 \frac{dy}{y} A(y) B(z/y,\ln \frac{y^2Q^2R^2}{4\mu^2}) = \int_z^1 \frac{dy}{y} A(z/y) B(y,\ln \frac{(z/y)^2Q^2R^2}{4\mu^2}) \,,
\end{align}
where $\otimes$ is the usual Mellin / DGLAP convolution for processes like hadron fragmentation, whereas $\boxtimes$ is the modified convolution for the angular observables as inclusive jets. It is worth noting that $\otimes = \boxtimes$ when the $B$ does not involve an angular variable. As there is never a situation in which both $A$ and $B$ involve angular variables, chain of $\boxtimes$ convolutions will reduce to all $\otimes$ convolution or all but one $\boxtimes$ convolution. 

With this new notation, we are now able to represent the \eq{jetRG} as
\begin{align}
\label{eq:jetRG2}
 \frac{d J_i}{d \ln \mu^2} = P_{ji}\boxtimes J_j = \sum_{n=0}^\infty a_s^{n+1} P_{ji}^{(n)} \boxtimes J_j\,,
\end{align}
where we made the flavor indices explicit and repeated indices imply implicit summation. We also define $a_s = \alpha_s/(4\pi)$ for QCD and $a = (g^2 N_c^2)/(16\pi^2)$ for $\mathcal{N}=4$ SYM, and $P_{ji}^{(n)}$ is the corresponding timelike splitting kernel at $\mathcal{O}(\alpha_s^{n+1})$ order. As the inclusive jet functions are single logarithmic in nature, the renormalized jet function takes the following form
\begin{align}
\label{eq:Jren}
J_i\left(z,\ln\frac{Q^2 R^2}{4\mu^2},\mu \right) =& \delta(1-z) + a_s \left[J_i^{(1,0)} + \textcolor{blue}{J_i^{(1,1)}}L\right] + a_s^2 \left[J_i^{(2,0)} + \textcolor{blue}{J_i^{(2,1)}}L + \textcolor{blue}{J_i^{(2,2)}}\frac{L^2}{2}\right]\nn\\
&+a_s\epsilon \left[J_{i\epsilon}^{(1,0)}+\textcolor{blue}{J_{i\epsilon}^{(1,1)}}L + \textcolor{blue}{J_{i\epsilon}^{(1,2)}}\frac{L^2}{2}\right] + \mathcal{O}(a_s^3, a_s^2\epsilon)\,,
\end{align}
where $L = \ln [4\mu^2/(Q^2R^2)]$. The notations $J_i^{(m,n)}$ denote $z$-dependent functions with coefficients $a_s^m \frac{L^n}{n!}$. Note that the terms explicitly proportional to $a_s\,\epsilon$ are also kept, as they are important when considering the relation to the bare jet function as we will see. Note that we work in $d=4-2\epsilon$. Then the RG evolution given in \eq{jetRG2} completely fixes the logarithmically enhanced blue terms. By comparing the LHS and RHS of \eq{jetRG2}, they are derived as
\begin{align}
\label{eq:Jblue}
J_i^{(1,1)} =& \sum_j P_{ji}^{(0)}\,,\nn\\
J_i^{(2,2)} =& \sum_j \left[P_{ki}^{(0)} \otimes P_{jk}^{(0)} + \beta_0 P_{ji}^{(0)}\right]\,,\nn\\
J_i^{(2,1)} =& \sum_j \left[P_{ji}^{(1)} - 2P_{ki}^{(0)}\otimes (P_{jk}^{(0)} \textcolor{red}{\ln y})\right] + P_{ki}^{(0)} \otimes J_k^{(1,0)}+ \beta_0 J_i^{(1,0)} \,,\nn\\
J_{i\epsilon}^{(1,1)} =& J_i^{(1,0)}\,,\nn\\
J_{i\epsilon}^{(1,2)} =& J_i^{(1,1)}\,.
\end{align}
Note that we choose to express all our terms using the usual DGLAP convolution $\otimes$. This results in an additional term involving $\textcolor{red}{\ln y}$ due to $\boxtimes$ convolution, where $y$ is the convolution variable integrated over. That is, the term involving $\textcolor{red}{\ln y}$ in $J_i^{(2,1)}$ is due to the modified convolution relative to the hadron fragmentation case and dropping this additional term would correspond to the standard DGLAP RG predictions. 

The \eq{jetRG2} also implies the relation between the bare and renormalized inclusive jet function as
\begin{align}
\label{eq:Jbarerel}
J_i\left(z,\ln\frac{Q^2 R^2}{4\mu^2},\mu \right) = Z_{ij}\boxtimes J_j^{\rm bare} = Z_{ij}\otimes J_j^{\rm bare}\,,
\end{align}
where absence of $\mu^2$ dependence, and thus $Q^2R^2/4$ dependence, in the bare function reduces the $\boxtimes$ convolution into the usual DGLAP convolution. We also define the inverse of $Z_{ij}$ such that
\begin{align}
\label{eq:Zinv}
 Z_{ij}\boxtimes Z_{jk}^{-1}  = \delta_{ik} \delta(1-z) 
\end{align}
to all orders. Then from Eqs.~\eqref{eq:jetRG2} and~\eqref{eq:Jbarerel}, we derive
\begin{align}
\label{eq:ZP}
\frac{d}{d\ln\mu^2}Z_{ik}\boxtimes Z_{kj}^{-1} = P_{ji} = \sum_{n=0}^\infty a_s^{n+1} P_{ji}^{(n)}\,.
\end{align}
As clearly expressed in \eq{jetRG2}, inclusive jets are still described by the timelike splitting kernels $P_{ji}$, albeit with the convolution structure modified. From our knowledge of the timelike splitting kernels, we are then able to extract the renormalization counterterms $Z$ order-by-order from \eq{ZP}. The relevant ones to two-loop order are
\begin{align}
Z_{ij}^{(0)} =& \delta_{ij}\delta(1-x)\,,\qquad\qquad &(Z_{ij}^{-1})^{(0)} =& \delta_{ij}\delta(1-x)\,,\nn\\
Z_{ij}^{(1)} =& -\frac{1}{\epsilon} P_{ji}^{(0)}\,,\qquad\qquad &(Z_{ij}^{-1})^{(0)} =& \frac{1}{\epsilon} P_{ji}^{(0)}\,,\nn\\
 Z_{ij}^{(2)} =& -\frac{1}{2\epsilon}P_{ji}^{(1)} + \frac{1}{2\epsilon^2}P_{ki}^{(0)}\otimes P_{jk}^{(0)}+\beta_0 \frac{1}{2\epsilon^2}P_{ji}^{(0)}\,.
\end{align}
From the explicit expressions of the renormalized jet function given in \eq{Jren}, we can then derive the form of the bare jet function from the relation \eq{Jbarerel}. To two-loop order, they are given as
\begin{align}
\label{eq:Jbare1}
J^{\rm bare}_i (z)=& J_i\left(z,\ln\frac{Q^2 R^2}{4\mu^2},\mu \right) + a_s^2 P_{ki}^{(0)} \otimes \left[J_{k\epsilon}^{(1,0)}+\textcolor{blue}{J_{k\epsilon}^{(1,1)}}(L-2\textcolor{red}{\ln y}) + \textcolor{blue}{J_{k\epsilon}^{(1,2)}}\frac{(L-2\textcolor{red}{\ln y})^2}{2}\right]\nn\\
&+ \frac{a_s^2}{\epsilon}P_{ki}^{(0)}\otimes\left[J_k^{(1,0)}+\textcolor{blue}{J_k^{(1,1)}}(L-2\textcolor{red}{\ln y})\right]  + \frac{a_s^2}{2\epsilon^2} P_{ki}^{(0)}\otimes P_{jk}^{(0)} \nn\\
&+\sum_j\left[\frac{a_s}{\epsilon}P_{ji}^{(0)}+ \frac{a_s^2}{2\epsilon} P_{ji}^{(1)} - \beta_0 \frac{a_s^2}{2\epsilon^2} P_{ji}^{(0)}
\right]\,,
\end{align}
where we again choose to express all our terms with the usual DGLAP convolution $\otimes$ and express the $\textcolor{red}{\ln y}$ arising from the $\boxtimes$ convolution explicitly. Dropping these terms would correspond to the standard DGLAP predictions. Explicit expressions of the logarithmic enhanced terms in blue were given in \eq{Jblue}.

On the other hand, the bare jet function can also be expressed as
\begin{align}
\label{eq:Jbare2}
J^{\rm bare}_i (z) =& \delta(1-z) + Z_\alpha a_s \left(\frac{4\mu^2}{Q^2R^2}\right)^\epsilon \left[\frac{J_{i,-1}^{(1)}}{\epsilon} + J_{i,0}^{(1)}\right]+ Z_\alpha^2 a_s^2 \left(\frac{4\mu^2}{Q^2R^2}\right)^{2\epsilon} \left[\frac{J_{i,-2}^{(2)}}{\epsilon^2} +\frac{J_{i,-1}^{(2)}}{\epsilon} + J_{i,0}^{(2)}\right]\nn\\
=&\delta(1-z) + a_s \left(\frac{J_{i,-1}^{(1)}}{\epsilon} +J_{i,0}^{(1)} + J_{i,-1}^{(1)} L\right)\nn\\
&+a_s^2 \left[\frac{J_{i,-2}^{(2)} - \beta_0 J_{i,-1}^{(1)}}{\epsilon^2} + \frac{(J_{i,-1}^{(2)} -\beta_0 J_{i,0}^{(1)}) + (2J_{i,-2}^{(2)} - \beta_0 J_{i,-1}^{(1)})L}{\epsilon} \right.\nn\\
&\left.\qquad+ J_{i,0}^{(2)}  + (2 J_{i,-1}^{(2)}  - \beta_0 J_{i,0}^{(1)} )L + (4 J_{i,-2}^{(2)}  - \beta_0 J_{i,-1}^{(1)} )\frac{L^2}{2}\right]\,,
\end{align}
where $Z_\alpha = 1- a_s \beta_0/\epsilon + \mathcal{O}(a_s^2)$ is the renormalization counterterms for the coupling renormalization. Here, terms are organized directly by coefficients of different $\epsilon^{-n}$ expansion, which is a natural way to organize the numerical calculations. In particular, terms associated with the divergent parts are directly associated with the inclusive jet RG given in \eq{jetRG2}. That is, comparing \eq{Jbare1} and \eq{Jbare2}, we find
\begin{align}
\label{eq:Jpred}
J_{i,-1}^{(1)}(z) &=\sum_j P_{ji}^{(0)}\,, \nn\\
J_{i,-2}^{(2)}(z) &=\frac{1}{2}\sum_j \left[P_{ki}^{(0)}\otimes P_{jk}^{(0)}+\beta_0P_{ji}^{(0)}\right] \,,\nn\\
J_{i,-1}^{(2)}(z) &=  \sum_j \left[\frac{1}{2}P_{ji}^{(1)} - 2 P_{ki}^{(0)} \otimes (P_{jk}^{(0)}\textcolor{red}{\ln y})\right] + P_{ki}^{(0)}\otimes J_k^{(1,0)} + \beta_0 J_i^{(1,0)} \,.
\end{align}
This represents the explicit predictions of our modified convolution structure at two-loop. This is in agreement with what was noted in~\cite{vanBeekveld:2024jnx}, and we derived it here from the RG analysis of our all-order factorization for inclusive jets presented in sec.~\ref{sec:fact}.

It is also interesting to consider the behavior for moments. As discussed extensively in sec.~\ref{sec:fact}, $N$-th moments of inclusive jet functions have identical RG structure as that of $N$ point projected energy correlators. Taking $N$-th moment of \eq{Jpred}, we find
\begin{align}
\label{eq:Jpredmom}
J_{i,-1}^{(1)}(N)\equiv \int_0^1 dz\, z^{N} J_{i,-1}^{(1)}(z) =&-\sum_j \gamma_{ji,T}^{(0)}(N+1)\,, \nn\\
J_{i,-2}^{(2)}(N)\equiv \int_0^1 dz\, z^{N} J_{i,-2}^{(2)}(z) =&\frac{1}{2}\sum_j \left[\gamma_{ki,T}^{(0)}(N+1) \gamma_{jk,T}^{(0)}(N+1) -\beta_0 \gamma_{ji,T}^{(0)}(N+1)\right] \,,\nn\\
J_{i,-1}^{(2)}(N)\equiv \int_0^1 dz\, z^{N} J_{i,-1}^{(2)}(z) =& \sum_j \left[-\frac{1}{2}\gamma_{ji,T}^{(1)}(N+1)  - 2 \dot{\gamma}_{ki,T}^{(0)}(N+1)\gamma_{jk,T}^{(0)}(N+1) \right] \nn\\
&- \gamma_{ki,T}^{(0)}(N+1) J_k^{(1,0)}(N) + \beta_0 J_i^{(1,0)}(N) \,,
\end{align}
where
\begin{align}
\label{eq:gammaQCD}
\gamma_{ji,T}^{(m)}(k)\equiv&-\int_0^1 \mathrm{~d}x\,x^{k-1} P_{ji}^{(m)}(x)\,,\nn\\
\dot{\gamma}_{ji,T}^{(m)}(k)\equiv&-\int_0^1 \mathrm{~d}x\,x^{k-1} \ln x\, P_{ji}^{(m)}(x) =-\partial_k \int_0^1 \mathrm{~d}x\,x^{k-1}\, P_{ji}^{(m)}(x) = \partial_k \gamma_{ji,T}^{(m)}(k)  \,,\nn\\
J_i^{(1,0)}(k) \equiv& \int_0^1 \mathrm{~d}x\,x^{k} J_i^{(1,0)}(x)\,.
\end{align}
Relative to the standard DGLAP expectation, $\dot{\gamma}_{ji,T}$ represents the additional term arising from the modified convolution structure. They are also exactly the results one expect for $N+1$ point projected energy correlators. We've also confirmed that our logarithmic terms match that of two-point and three-point energy correlators computed at the two-loop~\cite{Chen:2023zlx,Dixon:2019uzg}. 

Finally, the analogous expressions for $\mathcal{N}=4$ can be easily extracted from their universal splitting function and dropping terms proportional to $\beta_0$ in \eq{Jpredmom} as (also the expansion is now in $a \equiv \frac{g^2N_c}{16\pi^2}$)
\begin{align}
\label{eq:RGmomN4}
J_{-1,\mathcal{N}=4}^{(1)}(N)\equiv \int_0^1 dz\, z^{N} J_{-1,\mathcal{N}=4}^{(1)}(z) =&-\gamma_{T,\mathcal{N}=4}^{(0)}(N-1)\,, \nn\\
J_{-2,\mathcal{N}=4}^{(2)}(N)\equiv \int_0^1 dz\, z^{N} J_{-2,\mathcal{N}=4}^{(2)}(z) =&\frac{1}{2} \left[\gamma_{T,\mathcal{N}=4}^{(0)}(N-1) \gamma^{(0)}(N-1) -\beta_0 \gamma_{T,\mathcal{N}=4}^{(0)}(N-1)\right] \,,\nn\\
J_{-1,\mathcal{N}=4}^{(2)}(N)\equiv \int_0^1 dz\, z^{N} J_{-1,\mathcal{N}=4}^{(2)}(z) =& -\frac{1}{2}\gamma_{T,\mathcal{N}=4}^{(1)}(N-1)  - 2 \dot{\gamma}_{T,\mathcal{N}=4}^{(0)}(N-1)\gamma_{T,\mathcal{N}=4}^{(0)}(N-1)  \nn\\
&- \gamma_{T,\mathcal{N}=4}^{(0)}(N-1) J_{\mathcal{N}=4}^{(1,0)}(N) + \beta_0 J_{\mathcal{N}=4}^{(1,0)}(N) \,,
\end{align}
where
\begin{align}
\gamma_{T,\mathcal{N}=4}^{(L)}(k-2)\equiv&-\int_0^1 \mathrm{~d}x\,x^{k-1}\, P_{T,\rm{uni}}^{(m)}(x)\,,\nn\\
\dot{\gamma}_{T,\mathcal{N}=4}^{(L)}(k-2)\equiv&-\int_0^1 \mathrm{~d}x\,x^{k-1} \ln x\, P_{T,\rm{uni}}^{(m)}(x) =-\partial_k \int_0^1 \mathrm{~d}x\,x^{k-1}\, P_{T,\rm{uni}}^{(m)}(x) = \partial_k \gamma_{T,\mathcal{N}=4}^{(m)}(k-2)  \,,\nn\\
J_{i,\mathcal{N}=4}^{(1,0)}(k) \equiv& \int_0^1 \mathrm{~d}x\,x^{k} J_{i,\mathcal{N}=4}^{(1,0)}(x)\,.
\end{align}
Note that the arguments are shifted by two for the timelike anomalous dimension for $\mathcal{N}=4$ relative to the convention in \eq{gammaQCD}. This standard convention comes from the Mellin moments being shifted by two units when sum is performed
$\sum_j \gamma_{j \phi}(N)=\sum_j \gamma_{j \lambda}(N)=\sum_j \gamma_{j g}(N)=\gamma_{T,\mathcal{N}=4}(N-2)$. As discussed in sec.~\ref{sec:fact}, this RG expectation implies that the jet radius dependence $R^2$ for $N+1$ moment of the inclusive jets then has a power-law scaling governed by the \emph{spacelike} anomalous dimension of spin $N$.
\section{Two-Loop Calculation of the Inclusive Jet Function}\label{sec:calc}

To verify the factorization theorem derived in the sec.~\ref{sec:fact}, we now perform an explicit two-loop calculation for the moments of the inclusive jet function
\begin{equation}
	J_l^{\mathrm{bare}}(N)=\int_0^1 dz\, z^N J_l^{\mathrm{bare}}(z)\,,
\end{equation}
where $l = q, g$ corresponds to the quark and gluon jets in QCD. Additionally, we will compute the corresponding result for $\mathcal{N}=4$ SYM, denoted as $J_{\mathcal{N}=4}^{\mathrm{bare}}(N)$.

At two loops, the calculation decomposes into virtual corrections from NLO $1 \to 2$ splittings and real corrections from tree-level $1 \to 3$ splittings, expressed for both full $z$ dependence and moments as
\begin{align}
J_l^{\mathrm{bare},(2)}(z) =& J_l^{\mathrm{virtual},(2)}(z)+J_l^{\mathrm{real},(2)}(z)\,,\nonumber \\
	J_l^{\mathrm{bare},(2)}(N)=&J_l^{\mathrm{virtual},(2)}(N)+J_l^{\mathrm{real},(2)}(N)\,.
\end{align}
We will begin by explaining how these virtual and real corrections are computed. We will then compare our computation with the predicted poles of the moments of the bare jet functions, given in~\eq{Jpredmom}, based on the inclusive jet RG equation with modified convolution. Our results show perfect agreement and clearly reveal a deviation from the predictions of the standard timelike DGLAP equation. Finally, we will also present the two-loop jet functions relevant for exclusive jet production and compare them with the recent computation of the two-loop exclusive quark jet function presented in~\cite{Liu:2021xzi}.

\subsection{Virtual Corrections}

The virtual corrections for our two-loop computation can be expressed as
\begin{equation}
	J_l^{\rm{virtual},(2)}(z)\equiv \frac{1}{2}\sum_{ij}\int d\Phi_2^{c}(s,x)\,\sigma_{2,l\to ij}^{(1)}(s,x)\, \mathcal{M}_2(z,x)\,,
\end{equation}
where the overall $1/2$ is the symmetry factor for identical particles when  $i=j$ and gets canceled by permuting the indices when $i\neq j$. The notations $d\Phi_2^c$, $\sigma_{2,l\to ij}^{(1)}$, and $\mathcal{M}_2$ represent the two-particle collinear phase space, the two-particle one-loop collinear matrix element, and the two-particle measurement function, respectively. We will now explain each of these components in detail.

First, the two-particle collinear phase space $d\Phi_2^c$ is given by~\cite{Giele:1991vf,Ritzmann:2014mka}
\begin{equation}
	d\Phi_2^{c}(s,x)=\df s \df x \frac{[x(1-x)s]^{-\epsilon}}{(4\pi)^{2-\epsilon}\Gamma(1-\epsilon)}\,,
\end{equation}
where $s\geq 0$ represents the timelike virtuality of the initial parton, and  $x, 1-x$ are the momentum fractions of the two final-state partons. 

Next, the two-particle one-loop collinear matrix elements $\sigma_{2,l\to ij}^{(1)}$ are related to the $1\to2$ NLO splitting functions via 
\begin{align}
\sigma_{2,l\to ij}^{(1)}(s,x)=\left(\frac{\mu^2 e^{\gamma_E}}{4\pi}\right)^\epsilon \frac{2g^2}{s}P_{l\to ij}^{(1)}(x)\,.
\end{align}
For $\mathcal{N}=4$, the $1\to 2$ NLO splitting function is known to all orders in dimensional regularization $\epsilon$ \cite{Anastasiou:2003kj,Bern:2004cz}:
\begin{equation}
	P_{\mathcal{N}=4}^{(1)}(x)=2\Re\left[r_S^{(1)}(x,s,\epsilon) P_{\mathcal{N}=4}^{(0)}(x)\right]\,,
\end{equation}
written in terms of the tree-level $1\to 2$ splitting function 
\begin{equation}
	P_{\mathcal{N}=4}^{(0)}(x)=2N_c\left(\frac{1}{x}+\frac{1}{1-x}\right)
\end{equation}
and the ratio between NLO and tree-level expressed as
\begin{align}
	r_S^{(1)}(x,s,\epsilon)&=\frac{g^2N_c}{(4\pi)^2}\frac{e^{\epsilon\gamma_E}}{\epsilon^2}\left(\frac{\mu^2}{-s}\right)^\epsilon \frac{\Gamma^2(1-\epsilon)\Gamma(1+\epsilon)}{\Gamma(1-2\epsilon)}\notag\\
	&\times\left[-\frac{\pi\epsilon}{\sin(\pi\epsilon)}\left(\frac{1-x}{x}\right)^\epsilon+2\sum_{k=0}^\infty \epsilon^{2k+1}\text{Li}_{2k+1}\left(\frac{-x}{1-x}\right)\right]\,.
\end{align}
For QCD, we need NLO virtual correction to the $q\to qg$ splitting function for quark jets and the $g\to qq$ and $g\to gg$ splitting functions for gluon jets. These results were calculated in~\cite{Kosower:1999rx,Sborlini:2013jba} and are nicely summarized in \cite{Chen:2022pdu}. For convenience, we present them in~\App{sec:pert}.

\begin{figure}
\centering
    \centering
     \includegraphics[scale=0.6]{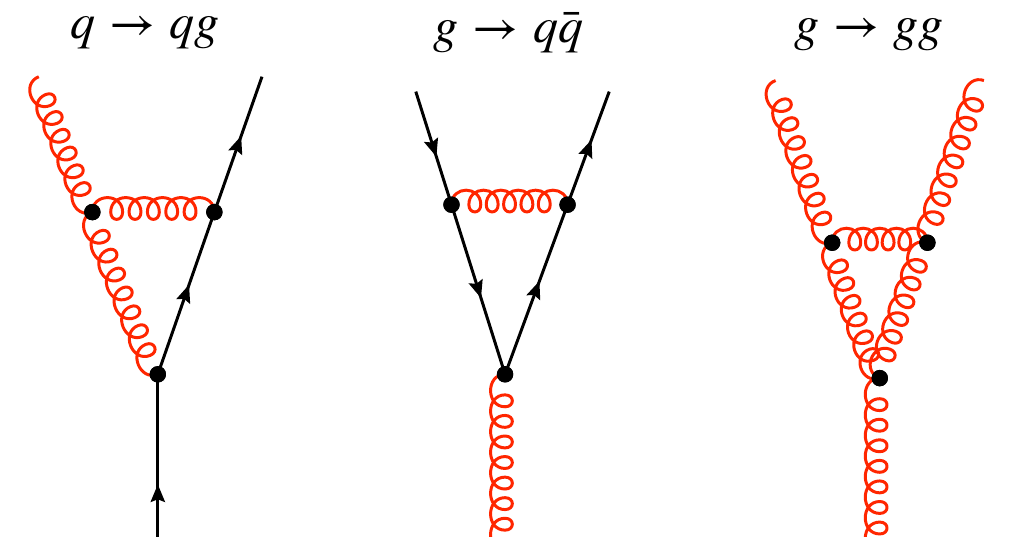}
     \vspace{1cm}
     \includegraphics[scale=0.6]{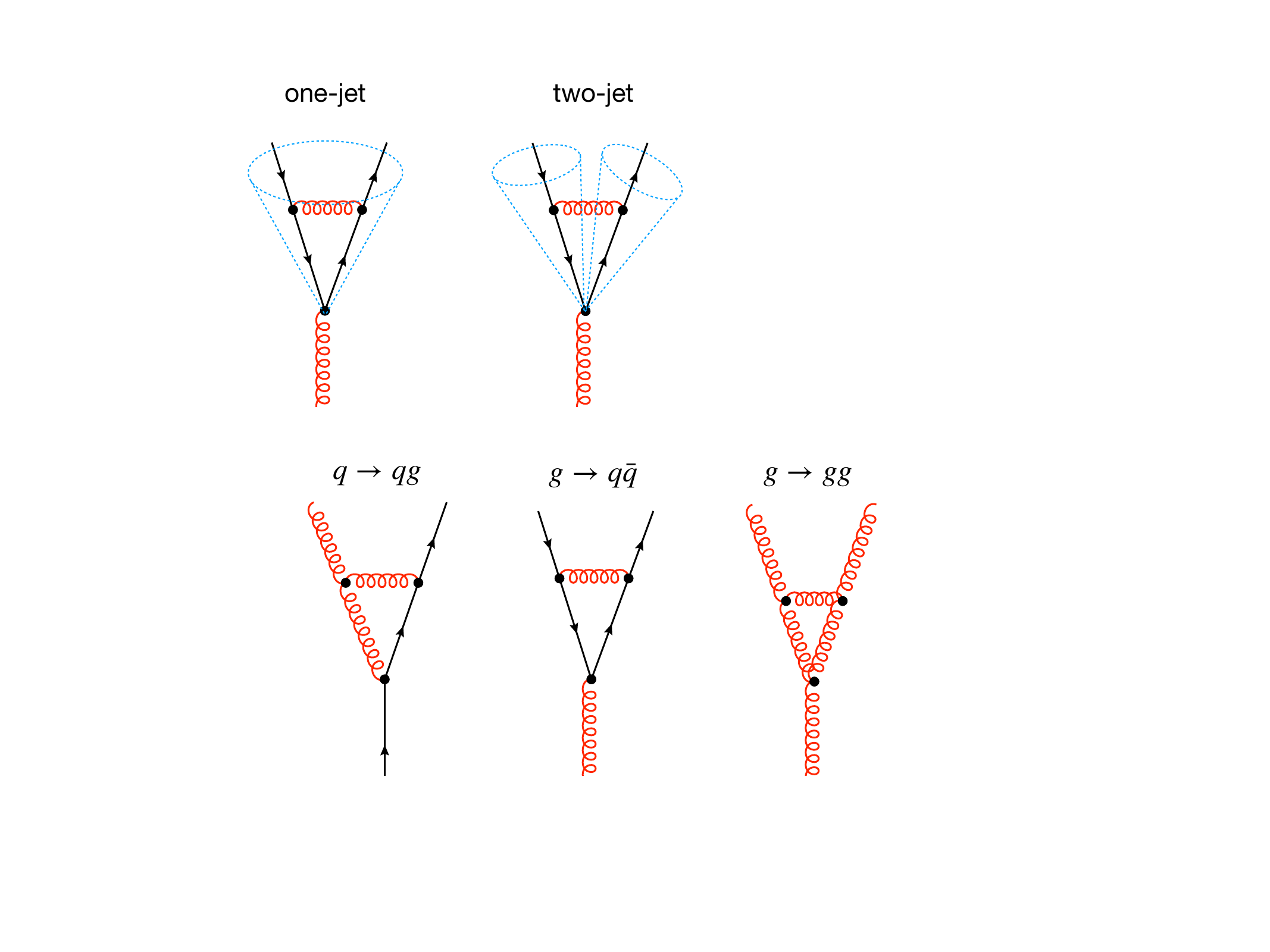}
        \caption{Upper panel: Examples of $1\to 2$ NLO splitting functions. Lower panel: Two possible jet configurations with two-particle final states.}
        \label{fig:virtual_config}
\end{figure}

Finally, the two-particle inclusive jet measurement functions have two possible configurations, corresponding to either one or two jets in the final state, as depicted in Fig.~\ref{fig:virtual_config}. For all $k_T$-type jet algorithms, the measurement function is given by~\cite{Kang:2016mcy}
\begin{equation}
\label{eq:meas2}
	\mathcal{M}_2(z,x)=\underbrace{\theta(\tilde s< 1)\delta(1-z)}_{\text{one-jet}}+\underbrace{ \theta(\tilde s>1)\left[\delta(x-z)+\delta(1-x-z)\right] }_{\text{two-jet}}\,,
\end{equation} 
where $\tilde{s}=s/(x(1-x)(E R)^2)$ with $E$ representing either the transverse momentum or energy of the initial parton, depending on whether we are considering the $p_T$ distribution or energy fraction ($E$ is denoted $Q/2$ in this section) of the jet. In the two-jet configuration, where two particles are clustered into separate jets, we still measure the transverse momentum or energy fraction $z$ once at a time as appropriate for single-inclusive jet measurements. The measurement function expression is identical to the one presented above in \eq{twoparticlemeas}, though with slightly different variables and notations. 

Using the $\delta$ function in the measurement function, we are able to represent the moments of the virtual corrections as 
\begin{align}
\label{eq:Jvirtmoments}
	J_i^{\text{virtual},(2)}(N)&=\int_0^1 \df z\, z^N \int d\Phi_2^{c}(s,x)\,\sigma_2^{(1)}(s,x)\, \mathcal{M}_2(z,x)\notag\\
&\hspace{-1cm}=\int d\Phi_2^{c}(s,x)\,\sigma_2^{(1)}(s,x)\,\theta(\tilde s< 1)+\int d\Phi_2^{c}(s,x)\,\sigma_2^{(1)}(s,x)\,\theta(\tilde s> 1)\left[x^N+(1-x)^N\right]\,.
\end{align}
The remaining $s,x$ integrals are also straightforward to evaluate analytically in \texttt{Mathematica}. For $N=1$ moment, we find zero virtual corrections in both $\mathcal{N}=4$ SYM and all channels in QCD. As the quantity of real interest is the sum of virtual and real corrections, i.e. the full bare jet function, we only present the $N=2,3,10$ moments in $\mathcal{N}=4$ SYM for illustration. They are given as
\begin{align}
\label{eq:n4_virtual_res}
	J_{\mathcal{N}=4}^{\text{virtual},(2)}(2)&=a^2 \left(\frac{4\mu^2}{Q^2 R^2}\right)^{2\epsilon} \left[\frac{4}{\epsilon^3}+\frac{32}{\epsilon^2}+\frac{240-68\zeta_2}{\epsilon}+\cdots\right]\notag\\
	J_{\mathcal{N}=4}^{\text{virtual},(2)}(3)&=a^2 \left(\frac{4\mu^2}{Q^2 R^2}\right)^{2\epsilon} \left[\frac{6}{2\epsilon^3}+\frac{48}{\epsilon^2}+\frac{360-102\zeta_2}{\epsilon}+\cdots\right]\notag\\
	J_{\mathcal{N}=4}^{\text{virtual},(2)}(10)&=a^2 \left(\frac{4\mu^2}{Q^2 R^2}\right)^{2\epsilon} \left[\frac{7129}{630\epsilon^3}+\frac{10019633}{99225\epsilon^2}+\frac{\frac{6617339483}{8334900}-\frac{121193}{630}\zeta_2}{\epsilon}+\cdots\right]
\end{align}
where we define $a=\frac{g^2 N_c}{16\pi^2}$ as in the previous section.

\subsection{Real Corrections}
The most complicated part of our calculation is the real emissions due to complicated integrals involving three-particle phase space. They are expressed as 
\begin{equation}
	J_l^{\text{real},(2)}(z)\equiv \frac{1}{6}\sum_{ijk}\int d\Phi_3^{c}(\{s\},\{x\})\,\sigma_{3,l\to ijk}^{(0)}(\{s\},\{x\})\, \mathcal{M}_3(z,\{s\},\{x\};\alpha)\,,
\end{equation}
where $d\Phi_3^c$, $\sigma_{3,l\to ijk}^{(0)}$, and $\mathcal{M}_3$ represent the three-particle collinear phase space, the three-particle tree-level collinear matrix element, and the three-particle measurement function, respectively. Similar to the virtual calculation, $1/6$ is the symmetry factor. 

The three-particle collinear phase space is given as~\cite{Ritzmann:2014mka}
\begin{equation}
	d\Phi_3^{c}(\{s\},\{x\}) = \df s_{ij}\, \df s_{ik}\, \df s_{jk}\,
	\df x_i\, \df x_j\, \df x_k\, \delta (1 - x_i - x_j - x_k) \frac{4
		\Theta (- \Delta_3^{\text{coll}}) (- \Delta_3^{\text{coll}})^{- \frac{1}{2}
			- \epsilon}}{(4 \pi)^{5 - 2 \epsilon} \Gamma (1 - 2 \epsilon)}\,,
\end{equation}
with $s_{jk}\geq 0$ the invariant mass of $j$, $k$ parton, $x_j\in [0,1]$ the energy fraction of $j$ parton, and the Gram determinant
\begin{equation}
	\Delta_3^{\text{coll}} = (x_k s_{ij} - x_i s_{jk} - x_j s_{ik})^2 - 4 x_i
	x_j s_{ik} s_{jk}\,.
\end{equation}
We use the shorthand notation $\{s\}=\{s_{ij},s_{ik},s_{jk}\}$ and $\{x\}=\{x_i,x_j,x_k\}$ to denote dependence in multiple invariant masses and energy fractions. 

The three-particle tree-level collinear matrix elements are related to the $1\to3$ tree-level splitting functions as 
\begin{align}
\sigma_{3,l\to ijk}^{(0)}=\left(\frac{\mu^2 e^{\gamma_e}}{4\pi}\right)^{2\epsilon} \frac{4g^4}{s_{jkl}^2} P_{l\to ijk}^{(0)}\,,
\end{align}
where $s_{ijk} = s_{ij} + s_{ik} + s_{jk}$. For $\mathcal{N}=4$, the $1\to 3$ tree-level splitting function is given as
\begin{equation}
	P_{\mathcal{N}=4}^{(0)} (\{s\},\{x\}) = N_c^2 \left[ \frac{s_{ijk}^2}{2 s_{ik}
		s_{jk}} \left( \frac{1}{x_i x_j} + \frac{1}{(1 - x_i) (1 - x_j)} \right) +
	\frac{s_{ijk}}{s_{ij} x_k} \left( \frac{1}{x_i} + \frac{1}{1 - x_i} \right)
	+ \text{perms} \right]
\end{equation}
with 5 additional permutations of the final-state gluons. For QCD, we have many more partonic channels compared to the virtual corrections discussed. For quark jets, we have the non-identical quark channel $q\to \bar q^\prime q^\prime q$, the identical quark channel $q\to \bar q q q$ and the gluon channel $q\to g g q$. For gluon jets, we have both the $n_f$ channel $g\to g q\bar q$ and the three-gluon channel $g\to ggg$. These results were calculated in \cite{Campbell:1997hg,Catani:1998nv} and we present them for completeness in \App{sec:pert}.

The most non-trivial part of the calculation is the three-particle measurement function denoted as $\mathcal{M}_3(z,\{s\},\{x\};\alpha)$. Here we specialize to the $k_T$-type jet algorithms, where $\alpha=0,-1,1$ correspond to Cambridge-Aachen, $k_T$, and anti-$k_T$ algorithms, respectively. Since we only present the poles in this paper, the results should be identical for all $\alpha$ if the proposed factorization were to hold for all the $k_T$-type jet algorithms. We confirm this to be the case within some numerical accuracy, but only present explicit results for the $\alpha=1$ case as the anti-$k_T$ is most phenomenologically relevant. 

\begin{figure}[!htbp]
\begin{center}
\includegraphics[scale=0.5]{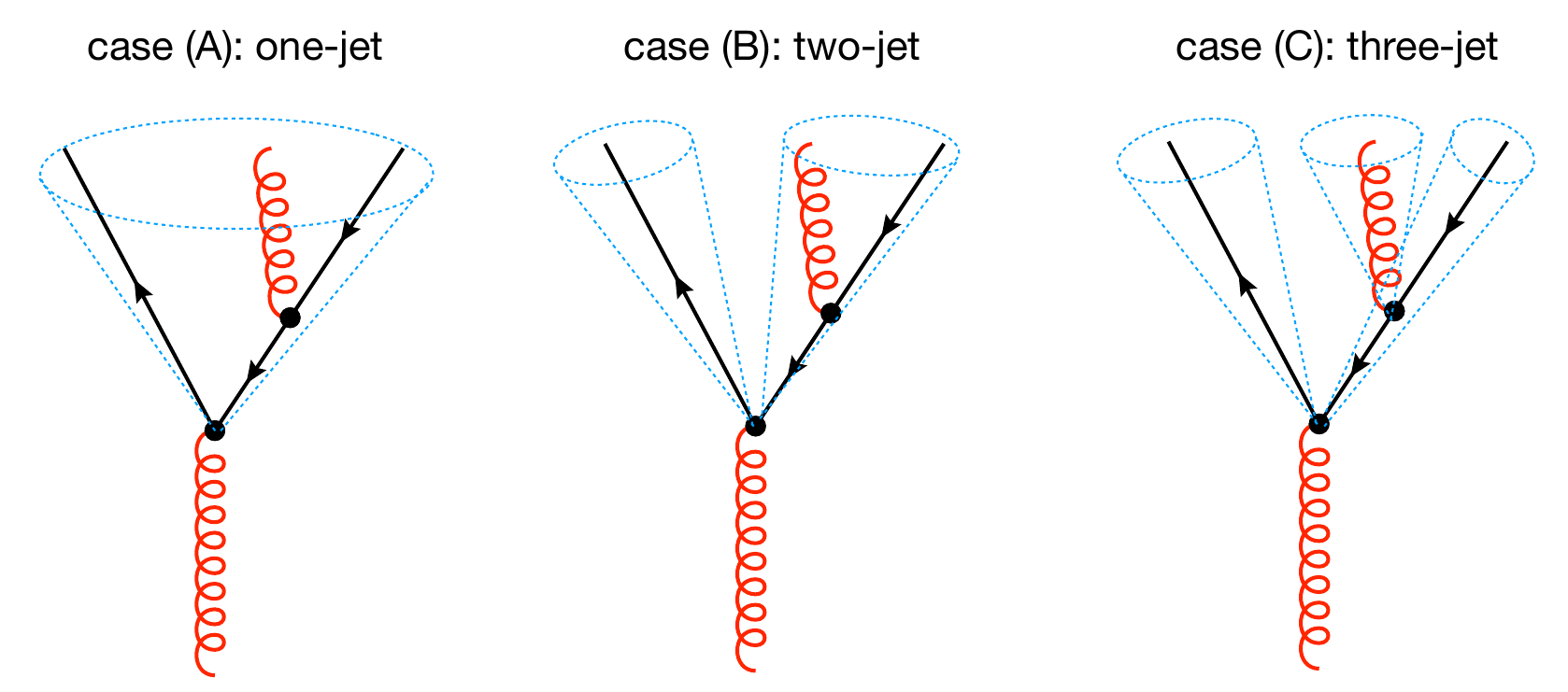} 
\end{center}
\caption{The three possibilities for clustering in the two-loop perturbative calculation of the jet function. In Case (A) all three partons are clustered into the same jet, in Case (B) two partons are clustered into one jet and one into another, in Case (C) all three partons are clustered into separate jets.}
\label{fig:ft}
\end{figure}

The measurement functions can be categorized into three different topologies, which we denote as (A), (B), and (C). In case (A), all three partons are clustered into the same jet. In case (B), two partons are clustered into one jet and one into another. Finally, in case (C), all three partons are clustered separately into a jet. These three configurations are shown in Fig.~\ref{fig:ft}. 

The phase space constraints for case (A), (B), (C) are derived from the definition of the $k_T$-type jet algorithm. It is convenient to discuss them with a definite energy ordering between the final-state partons. Without loss of generality, we assume an energy ordering of the partons, $\theta_{ijk}\equiv \theta(x_i\leq x_j\leq x_k) $ to derive measurement functions with definite energy ordering. Similar to the virtual corrections above, we introduce the angular variable $\tilde{s}_{jk}=s/(x_j x_k (E R)^2)$ with $E$ representing either the transverse momentum or energy of the initial parton, depending on whether we are considering the $p_T$ distribution or energy fraction of the jet. 

For case (A), we find
\begin{align}
\label{eq:measA}
	\mathcal{M}_{3,ijk}^{\mathrm{AI}}&=\theta\left(\sij<\zjk \sjk\right)\theta\left(\sij<\zjk \sik\right)\nn\\
 &\hspace{5cm}\times\theta\left(\frac{x_i}{x_j}\sik+\sjk<1+\frac{x_i}{x_j}+\frac{x_i}{x_i+x_j}\sij\right)\theta_{ijk}\nn\\
	\mathcal{M}_{3,ijk}^{\mathrm{AII}}&=\left[1-\theta\left(\sij\leq\zjk \sik\right)\right]\theta\left(\sik<1\right)\theta\left(\sik\leq\sjk\right)\nn\\
 &\hspace{5cm}\times\theta\left(\frac{x_i}{x_k}\sij+\sjk<1+\frac{x_i}{x_k}+\frac{x_i}{x_i+x_k}\sik\right)\theta_{ijk}\nn\\
	\mathcal{M}_{3,ijk}^{\mathrm{AIII}}&=\left[1-\theta\left(\sij\leq\zjk \sjk\right)\right]\theta\left(\sjk<1\right)\theta\left(\sjk\leq\sik\right)\nn\\
 &\hspace{5cm}\times\theta\left(\frac{x_j}{x_k}\sij+\sik<1+\frac{x_j}{x_k}+\frac{x_j}{x_j+x_k}\sjk\right)\theta_{ijk}
\end{align}

where the different terms $\mathrm{AI},\mathrm{AII},\mathrm{AIII}$ respectively correspond to the cases where the pair  $[i,j]$, $[i,k]$, and $[j,k]$ cluster first. After the first clustering of the pair of partons, the third parton then also is clustered together to form a single jet. The subscript $ijk$ denotes the energy ordering $\theta_{ijk}$. For case (B), we find

\begin{align}
	\mathcal{M}_{3,ijk}^{\mathrm{BIa}}&=\theta\left(\sij<\zjk \sjk\right)\theta\left(\sij<\zjk\right)\theta\left(\sij<\zjk\sik\right)\nn\\
	&\hspace{5cm}\times\theta\left(\frac{x_i}{x_j}\sik+\sjk\geq 1+\frac{x_i}{x_j}+\frac{x_i}{x_i+x_j}\sij\right)\theta_{ijk}\nn\\	\mathcal{M}_{3,ijk}^{\mathrm{BIb}}&=\theta\left(\sij>\zjk\right)\theta\left(\sjk>1\right)\theta\left(\sik>1\right)\theta\left(\sij<1\right)\nn\\
	\mathcal{M}_{3,ijk}^{\mathrm{BII}}&=\left[1-\theta\left(\sij\leq\zjk \sik\right)\right]\theta\left(\sik<1\right)\theta\left(\sik<\sjk\right)\nn\\
	&\hspace{5cm}\times\theta\left(\frac{x_i}{x_k}\sij+\sjk\geq 1+\frac{x_i}{x_k}+\frac{x_i}{x_i+x_k}\sik\right)\theta_{ijk}\nn\\
	\mathcal{M}_{3,ijk}^{\mathrm{BIII}}&=\left[1-\theta\left(\sij \leq \zjk \sjk\right)\right]\theta\left(\sjk<1\right)\theta\left(\sjk<\sik\right)\nn\\
	&\hspace{5cm}\times \theta\left(\frac{x_j}{x_k}\sij+\sik\geq 1+\frac{x_j}{x_k}+\frac{x_j}{x_j+x_k}\sjk\right)\theta_{ijk}
\end{align}

where $\mathrm{BII}$ and $\mathrm{BIII}$ respectively correspond to the cases where the pair  $[i,k]$ and $[j,k]$ cluster first. Unlike case (A), however, after the first clustering of the pair of partons, the third parton becomes a separate jet. The cases $\mathrm{BIa}$ and $\mathrm{BIb}$ correspond to the case where the pair $[i,j]$ and single parton $[k]$ are separately identified as a jet. For the case $\mathrm{BIa}$, the pair $[i,j]$ clusters into a jet first, whereas for the case $\mathrm{BIb}$, a single parton $[k]$ becomes a jet first. Finally, the measurement function of the case (C) can be written as
\begin{align}
\mathcal{M}_{3,ijk}^{\mathrm{C}}=& \theta_{ijk}\theta(\sij\geq 1)\theta(\sik>1)\theta(\sjk>1)\,,
\end{align}
which describes the scenario where three particles are each clustered into a separate jet.

Combining all these configurations, the full three-particle $k_T$-type jet measurement with all energy ordering can be written as
\begin{align}
	&\mathcal{M}_3(z,\{s\},\{x\};\alpha)=\delta(1-z)\left[\mathcal{M}_{3,ijk}^{\mathrm{A}}+\mathcal{M}_{3,ikj}^{\mathrm{A}}+\mathcal{M}_{3,jik}^{\mathrm{A}}+\mathcal{M}_{3,jki}^{\mathrm{A}}+\mathcal{M}_{3,kij}^{\mathrm{A}}+\mathcal{M}_{3,kji}^{\mathrm{A}}\right]\notag\\
	&+[\delta(z-x_i-x_j)+\delta(z-x_l)]\left[\mathcal{M}_{3,ijk}^{\mathrm{BI}}+\mathcal{M}_{3,ikj}^{\mathrm{BII}}+\mathcal{M}_{3,jik}^{\mathrm{BI}}+\mathcal{M}_{3,jki}^{\mathrm{BII}}+\mathcal{M}_{3,kij}^{\mathrm{BIII}}+\mathcal{M}_{3,kji}^{\mathrm{BIII}}\right]\notag\\
	&+[\delta(z-x_i-x_k)+\delta(z-x_k)]\left[\mathcal{M}_{3,ijk}^{\mathrm{BII}}+\mathcal{M}_{3,ikj}^{\mathrm{BI}}+\mathcal{M}_{3,jik}^{\mathrm{BIII}}+\mathcal{M}_{3,jki}^{\mathrm{BIII}}+\mathcal{M}_{3,kij}^{\mathrm{BI}}+\mathcal{M}_{3,kji}^{\mathrm{BII}}\right]\notag\\
	&+[\delta(z-x_j-x_k)+\delta(z-x_j)]\left[\mathcal{M}_{3,ijk}^{\mathrm{BIII}}+\mathcal{M}_{3,ikj}^{\mathrm{BIII}}+\mathcal{M}_{3,jik}^{\mathrm{BII}}+\mathcal{M}_{3,jki}^{\mathrm{BI}}+\mathcal{M}_{3,kij}^{\mathrm{BII}}+\mathcal{M}_{3,kji}^{\mathrm{BI}}\right]\notag\\
	&+\left[\delta(z-x_i)+\delta(z-x_j)+\delta(z-x_k)\right]\left[\mathcal{M}_{3,ijk}^{\mathrm{C}}+\mathcal{M}_{3,ikj}^{\mathrm{C}}+\mathcal{M}_{3,jik}^{\mathrm{C}}+\mathcal{M}_{3,jki}^{\mathrm{C}}+\mathcal{M}_{3,kij}^{\mathrm{C}}+\mathcal{M}_{3,kji}^{\mathrm{C}}\right]\,,
\end{align}
where the delta functions denote the measurement of the jet momentum or energy fraction with respect to the initial fragmenting parton. Again, if we take the moments to compute $J_i^{\text{real},(2)}(N)$, similar to the virtual case in \eq{Jvirtmoments}, the integration with respect to $z$ will turn the $\delta$ functions into some powers, e.g. $\delta(z-x_j-x_k) \to (x_j+x_k)^N$.

To perform the numerical calculation, we 
parameterize to disentangle all possible soft and collinear divergences from each other with $\epsilon$-expansion and proper plus distributions. This gives rise to a five-fold integral which is then implemented in the Fortran version of the \texttt{Cuba} library~\cite{Hahn:2004fe}. We sampled 90 million points for each moment. In this paper, we also only compute the pole coefficients as they already allow us to verify the renormalization group structure. The exact parameterization, constant values of $\mathcal{O}(\epsilon^0)$ for different $k_T$-type algorithms, and how we implement these numerics will be presented in a forthcoming paper.

For illustration, we also present the $N=1,2,3,10$ moments for of the real corrections for $\mathcal{N}=4$ SYM. They are given as
\begin{align}\label{eq:n4_real_res}
	J_{\mathcal{N}=4}^{\text{real},(2)}(1)&= a^2 \left(\frac{4\mu^2}{Q^2 R^2}\right)^{2\epsilon}\left[-\frac{3.60739\times 10^{-8}}{\epsilon^4}+\frac{0.00210548}{\epsilon^3}+\frac{0.046837}{\epsilon^2}+\frac{0.296656}{\epsilon}+\cdots\right]\notag\\
	J_{\mathcal{N}=4}^{\text{real},(2)}(2)&=  a^2 \left(\frac{4\mu^2}{Q^2 R^2}\right)^{2\epsilon}\left[\frac{5.89426\times 10^{-8}}{\epsilon^4}-\frac{3.99816}{\epsilon^3}-\frac{23.9439}{\epsilon^2}-\frac{66.2994}{\epsilon}+\cdots\right]\notag\\
	J_{\mathcal{N}=4}^{\text{real},(2)}(3)&=  a^2 \left(\frac{4\mu^2}{Q^2 R^2}\right)^{2\epsilon}\left[\frac{1.06562\times 10^{-7}}{\epsilon^4}-\frac{5.99813}{\epsilon^3}-\frac{29.9355}{\epsilon^2}-\frac{45.0032}{\epsilon}+\cdots\right]\notag\\
	J_{\mathcal{N}=4}^{\text{real},(2)}(10)&= a^2 \left(\frac{4\mu^2}{Q^2 R^2}\right)^{2\epsilon}\left[\frac{4.38956\times 10^{-7}}{\epsilon^4}-\frac{11.3137}{\epsilon^3}-\frac{36.8837}{\epsilon^2}+\frac{113.569}{\epsilon}+\cdots\right]
\end{align}

\subsection{Results for $\mathcal{N}=4$ Super Yang-Mills}\label{sec:n4}

In this subsection, we summarize the two-loop results for $\mathcal{N}=4$ SYM and compare them with the RG predictions given in sec.~\ref{sec:renorm}. To obtain the two-loop bare jet function, we need to combine the virtual correction in Eq.~\eqref{eq:n4_virtual_res} and the real emissions in Eq.~\eqref{eq:n4_real_res}. We presented the separate contributions for the virtual and real corrections separately for the $N=1,2,3,10$ moments above. The explicit expressions for the first $10$ moments read
\begin{align}
	J_{\mathcal{N}=4}^{\text{bare},(2)}(1)&=  a^2 \left(\frac{4\mu^2}{Q^2 R^2}\right)^{2\epsilon}\left[-\frac{3.60739\times 10^{-8}}{\epsilon^4}+\frac{0.00210548}{\epsilon^3}+\frac{0.046837}{\epsilon^2}+\frac{0.296656}{\epsilon}+\cdots\right]\notag\\
	J_{\mathcal{N}=4}^{\text{bare},(2)}(2)&=  a^2 \left(\frac{4\mu^2}{Q^2 R^2}\right)^{2\epsilon}\left[\frac{5.89426\times 10^{-8}}{\epsilon^4}+\frac{0.00184038}{\epsilon^3}+\frac{8.05613}{\epsilon^2}+\frac{61.845}{\epsilon}+\cdots\right]\notag\\
	J_{\mathcal{N}=4}^{\text{bare},(2)}(3)&= a^2 \left(\frac{4\mu^2}{Q^2 R^2}\right)^{2\epsilon}\left[\frac{1.06562\times 10^{-7}}{\epsilon^4}+\frac{0.00187143}{\epsilon^3}+\frac{18.0645}{\epsilon^2}+\frac{147.214}{\epsilon}+\cdots\right]\notag\\
	J_{\mathcal{N}=4}^{\text{bare},(2)}(4)&= a^2 \left(\frac{4\mu^2}{Q^2 R^2}\right)^{2\epsilon}\left[\frac{1.54135\times 10^{-7}}{\epsilon^4}+\frac{0.00194754}{\epsilon^3}+\frac{26.9545}{\epsilon^2}+\frac{225.766}{\epsilon}+\cdots\right]\notag\\
	J_{\mathcal{N}=4}^{\text{bare},(2)}(5)&=  a^2 \left(\frac{4\mu^2}{Q^2 R^2}\right)^{2\epsilon}\left[\frac{2.01459\times 10^{-7}}{\epsilon^4}+\frac{0.00175942}{\epsilon^3}+\frac{34.7894}{\epsilon^2}+\frac{297.621}{\epsilon}+\cdots\right]\notag\\
	J_{\mathcal{N}=4}^{\text{bare},(2)}(6)&=  a^2 \left(\frac{4\mu^2}{Q^2 R^2}\right)^{2\epsilon}\left[\frac{2.49041\times 10^{-7}}{\epsilon^4}+\frac{0.0017578}{\epsilon^3}+\frac{41.7759}{\epsilon^2}+\frac{363.986}{\epsilon}+\cdots\right]\notag\\
	J_{\mathcal{N}=4}^{\text{bare},(2)}(7)&=  a^2 \left(\frac{4\mu^2}{Q^2 R^2}\right)^{2\epsilon}\left[\frac{2.96476\times 10^{-7}}{\epsilon^4}+\frac{0.0018432}{\epsilon^3}+\frac{48.0908}{\epsilon^2}+\frac{425.893}{\epsilon}+\cdots\right]\notag\\
	J_{\mathcal{N}=4}^{\text{bare},(2)}(8)&= a^2 \left(\frac{4\mu^2}{Q^2 R^2}\right)^{2\epsilon}\left[\frac{3.44086\times 10^{-7}}{\epsilon^4}+\frac{0.00188192}{\epsilon^3}+\frac{53.8541}{\epsilon^2}+\frac{483.987}{\epsilon}+\cdots\right]\notag\\
	J_{\mathcal{N}=4}^{\text{bare},(2)}(9)&= a^2 \left(\frac{4\mu^2}{Q^2 R^2}\right)^{2\epsilon}\left[\frac{3.91669\times 10^{-7}}{\epsilon^4}+\frac{0.00185897}{\epsilon^3}+\frac{59.1646}{\epsilon^2}+\frac{538.901}{\epsilon}+\cdots\right]\notag\\
	J_{\mathcal{N}=4}^{\text{bare},(2)}(10)&=  a^2 \left(\frac{4\mu^2}{Q^2 R^2}\right)^{2\epsilon}\left[\frac{4.38956\times 10^{-7}}{\epsilon^4}+\frac{0.00214815}{\epsilon^3}+\frac{64.0952}{\epsilon^2}+\frac{591.064}{\epsilon}+\cdots\right]
\end{align}
As discussed in sec.~\ref{sec:renorm}, the $1/\epsilon^4$ and $1/\epsilon^3$ poles are expected to cancel when the virtual and real contributions are combined, following the inclusive jet RG structure. The fact that the small, nonzero numerical size of these quartic and cubic pole values are of a similar order across different moments suggests that these deviations from zero are due to numerical effects. On the other hand, according to the inclusive jet RG structure, the size of the quadratic and linear poles across different moments are given by $J_{-2,\mathcal{N}=4}^{(2)}(N)$ and $J_{-1,\mathcal{N}=4}^{(2)}(N)$, respectively, from~\eq{RGmomN4}. 

The first moment $N=1$ evaluates to zero, and thus our numerical values away from zero for $N=1$ establish the baseline for the expected numerical precision at higher moments. Indeed, we find that our $1/\epsilon^2$ poles exhibit a similar order of deviation from what is expected based on the inclusive jet RG result given in~\eq{RGmomN4}. However, since the quadratic poles are identical between the inclusive jet RG and the usual DGLAP, they are not sufficient to distinguish the two factorization structures. The size of the $1/\epsilon$ pole is of particular interest, as the inclusive jet RG and DGLAP predict different values. As discussed in sec.~\ref{sec:renorm}, the DGLAP prediction corresponds to dropping the derivative terms of the form $\dot{\gamma}_T$ from the inclusive jet RG. While our numerics are not yet fully optimized --- which we plan to improve, particularly to enhance the accuracy of the constant terms --- we can nonetheless confirm that our results are consistent with the inclusive jet RG and deviate from the naive expectations based on DGLAP evolution.

\begin{figure}[!htbp]
	\includegraphics[scale=0.6]{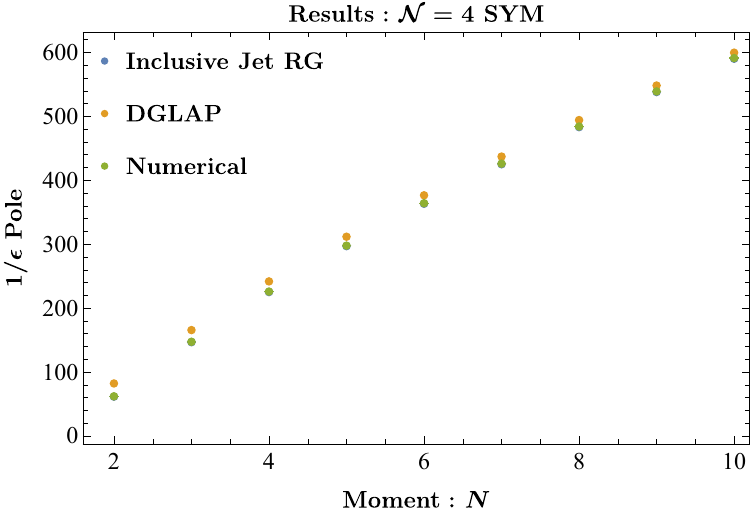}
	\includegraphics[scale=0.6]{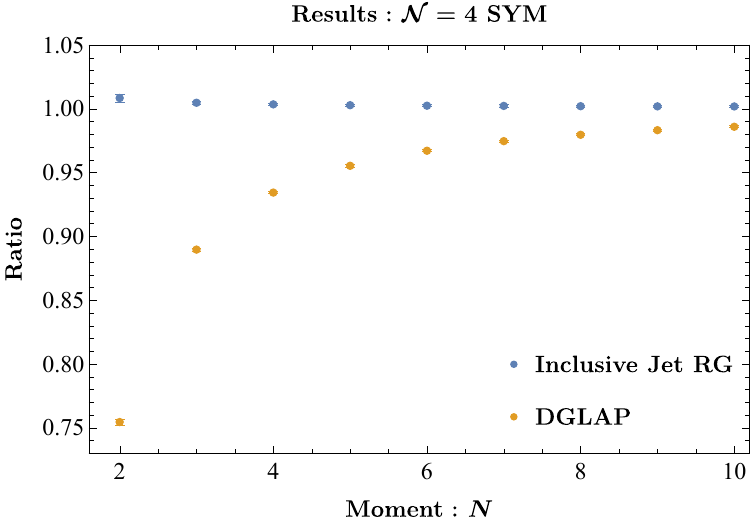}
	\caption{Comparison of Monte Carlo results for the $1/\epsilon$ linear pole in the two-loop bare $\mathcal{N}=4$ SYM jet function with the expected value derived from inclusive jet RG and DGLAP. In the left panel, we present the absolute values of the linear pole for inclusive jet RG, DGLAP, and our Monte Carlo results. In the right panel, we calculate the relative difference between the MC and what is expected from the RG by taking the ratio and find good agreement between the MC and the inclusive jet RG.}
	\label{fig:n4_plot}
\end{figure}

In Fig.~\ref{fig:n4_plot}, we compare the size of the linear pole for moments $N=2$ to $N=10$ and show that the Monte Carlo predictions align much more closely with the inclusive jet RG than with the DGLAP. Specifically, from the ratio defined as $\text{Numerical value}/\text{RG prediction}$, we find less than $1\%$ deviation with respect to the inclusive jet RG values and deviations ranging from $1.5\%\sim 25\%$ with respect to the DGLAP values, depending on the moment under consideration. This provides a highly nontrivial confirmation that our inclusive jet RG is correct for small-$R$ factorization.

\subsection{Results for QCD Jets}\label{sec:n4}

We also compute the two-loop bare jet function results for both quark jet and gluon jets. For convenience, we only present the total bare jet function, rather than present individaul contributions for virtual and real corrections as we have done for some of the moments for $\mathcal{N}=4$.

For quark jet, we find
\begin{align}
	J_q^{\text{bare},(2)}(1)&=Z_{\alpha}^2 a_s^2 \left(\frac{4\mu^2}{Q^2 R^2}\right)^{2\epsilon}\Bigg[C_F T_F n_f\left(\frac{4.29528\times 10^{-8}}{\epsilon^3}+\frac{0.000228517}{\epsilon^2}-\frac{0.000536371}{\epsilon}\right)\notag\\
	&+C_F^2\left(-\frac{1.48541\times 10^{-8}}{\epsilon^4}-\frac{0.000114897}{\epsilon^3}+\frac{0.0100901}{\epsilon^2}+\frac{0.131474}{\epsilon}\right)\notag\\
	&+C_F C_A\left(-\frac{4.13074\times 10^{-9}}{\epsilon^4}+\frac{0.000111077}{\epsilon^3}-\frac{0.0189275}{\epsilon^2}+\frac{0.0206227}{\epsilon}\right)\Bigg]\notag\\
	J_q^{\text{bare},(2)}(2)&=Z_{\alpha}^2 a_s^2 \left(\frac{4\mu^2}{Q^2 R^2}\right)^{2\epsilon}\Bigg[C_F T_F n_f\left(-\frac{1.57726\times 10^{-8}}{\epsilon^3}+\frac{1.76682}{\epsilon^2}+\frac{17.6214}{\epsilon}\right)\notag\\
	&+C_F^2\left(-\frac{4.49525\times 10^{-8}}{\epsilon^4}-\frac{0.000174055}{\epsilon^3}+\frac{6.25658}{\epsilon^2}+\frac{47.3899}{\epsilon}\right)\notag\\
	&+C_F C_A\left(\frac{1.58691\times 10^{-8}}{\epsilon^4}+\frac{0.0000339994}{\epsilon^3}-\frac{7.14685}{\epsilon^2}-\frac{65.4817}{\epsilon}\right)\Bigg]\notag\\
	J_q^{\text{bare},(2)}(3)&=Z_{\alpha}^2 a_s^2 \left(\frac{4\mu^2}{Q^2 R^2}\right)^{2\epsilon}\Bigg[C_F T_F n_f\left(-\frac{4.9401\times 10^{-8}}{\epsilon^3}+\frac{2.78015}{\epsilon^2}+\frac{27.5799}{\epsilon}\right)\notag\\
	&+C_F^2\left(-\frac{6.00801\times 10^{-8}}{\epsilon^4}+\frac{0.0000188765}{\epsilon^3}+\frac{11.7786}{\epsilon^2}+\frac{93.1792}{\epsilon}\right)\notag\\
	&+C_F C_A\left(\frac{2.58559\times 10^{-8}}{\epsilon^4}-\frac{0.0000356006}{\epsilon^3}-\frac{9.80104}{\epsilon^2}-\frac{90.0879}{\epsilon}\right)\Bigg]\notag\\
	J_q^{\text{bare},(2)}(4)&=Z_{\alpha}^2 a_s^2 \left(\frac{4\mu^2}{Q^2 R^2}\right)^{2\epsilon}\Bigg[C_F T_F n_f\left(-\frac{5.50888\times 10^{-8}}{\epsilon^3}+\frac{3.496}{\epsilon^2}+\frac{35.0271}{\epsilon}\right)\notag\\
	&+C_F^2\left(-\frac{7.49318\times 10^{-8}}{\epsilon^4}+\frac{0.0000514093}{\epsilon^3}+\frac{16.7863}{\epsilon^2}+\frac{137.448}{\epsilon}\right)\notag\\
	&+C_F C_A\left(\frac{3.5848\times 10^{-8}}{\epsilon^4}-\frac{0.000119016}{\epsilon^3}-\frac{11.5337}{\epsilon^2}-\frac{108.05}{\epsilon}\right)\Bigg]\notag\\
	J_q^{\text{bare},(2)}(5)&=a_s^2 \left(\frac{4\mu^2}{Q^2 R^2}\right)^{2\epsilon}\Bigg[C_F T_F n_f\left(\frac{4.42018\times 10^{-8}}{\epsilon^3}+\frac{4.05272}{\epsilon^2}+\frac{41.1655}{\epsilon}\right)\notag\\
	&+C_F^2\left(-\frac{8.99519\times 10^{-8}}{\epsilon^4}+\frac{0.000104789}{\epsilon^3}+\frac{21.3864}{\epsilon^2}+\frac{180.227}{\epsilon}\right)\notag\\
	&+C_F C_A\left(\frac{4.59458\times 10^{-8}}{\epsilon^4}-\frac{0.000162372}{\epsilon^3}-\frac{12.8643}{\epsilon^2}-\frac{123.028}{\epsilon}\right)\Bigg]\notag\\
	J_q^{\text{bare},(2)}(6)&=Z_{\alpha}^2 a_s^2 \left(\frac{4\mu^2}{Q^2 R^2}\right)^{2\epsilon}\Bigg[C_F T_F n_f\left(\frac{2.80696\times 10^{-7}}{\epsilon^3}+\frac{4.50966}{\epsilon^2}+\frac{46.4661}{\epsilon}\right)\notag\\
	&+C_F^2\left(-\frac{1.05026\times 10^{-7}}{\epsilon^4}-\frac{0.0000348}{\epsilon^3}+\frac{25.6407}{\epsilon^2}+\frac{221.403}{\epsilon}\right)\notag\\
	&+C_F C_A\left(\frac{5.59327\times 10^{-8}}{\epsilon^4}-\frac{0.000171823}{\epsilon^3}-\frac{13.9581}{\epsilon^2}-\frac{136.093}{\epsilon}\right)\Bigg]\notag\\
	J_q^{\text{bare},(2)}(7)&=Z_{\alpha}^2 a_s^2 \left(\frac{4\mu^2}{Q^2 R^2}\right)^{2\epsilon}\Bigg[C_F T_F n_f\left(\frac{1.97426\times 10^{-7}}{\epsilon^3}+\frac{4.89802}{\epsilon^2}+\frac{51.1715}{\epsilon}\right)\notag\\
	&+C_F^2\left(-\frac{1.2007\times 10^{-7}}{\epsilon^4}-\frac{0.00001216}{\epsilon^3}+\frac{29.6134}{\epsilon^2}+\frac{261.123}{\epsilon}\right)\notag\\
	&+C_F C_A\left(\frac{6.59248\times 10^{-8}}{\epsilon^4}-\frac{0.000199306}{\epsilon^3}-\frac{14.8926}{\epsilon^2}-\frac{147.759}{\epsilon}\right)\Bigg]\notag\\
	J_q^{\text{bare},(2)}(8)&=Z_{\alpha}^2 a_s^2 \left(\frac{4\mu^2}{Q^2 R^2}\right)^{2\epsilon}\Bigg[C_F T_F n_f\left(\frac{2.51072\times 10^{-7}}{\epsilon^3}+\frac{5.23592}{\epsilon^2}+\frac{55.4219}{\epsilon}\right)\notag\\
	&+C_F^2\left(-\frac{1.34922\times 10^{-7}}{\epsilon^4}-\frac{0.0000186693}{\epsilon^3}+\frac{33.3317}{\epsilon^2}+\frac{299.394}{\epsilon}\right)\notag\\
	&+C_F C_A\left(\frac{7.59119\times 10^{-8}}{\epsilon^4}-\frac{0.000174782}{\epsilon^3}-\frac{15.71}{\epsilon^2}-\frac{158.354}{\epsilon}\right)\Bigg]\notag\\
	J_q^{\text{bare},(2)}(9)&=Z_{\alpha}^2 a_s^2 \left(\frac{4\mu^2}{Q^2 R^2}\right)^{2\epsilon}\Bigg[C_F T_F n_f\left(\frac{1.23823\times 10^{-7}}{\epsilon^3}+\frac{5.53525}{\epsilon^2}+\frac{59.3125}{\epsilon}\right)\notag\\
	&+C_F^2\left(-\frac{1.49947\times 10^{-7}}{\epsilon^4}-\frac{0.0000271812}{\epsilon^3}+\frac{36.8418}{\epsilon^2}+\frac{336.492}{\epsilon}\right)\notag\\
	&+C_F C_A\left(\frac{8.59247\times 10^{-8}}{\epsilon^4}-\frac{0.000184862}{\epsilon^3}-\frac{16.4411}{\epsilon^2}-\frac{168.135}{\epsilon}\right)\Bigg]\notag\\
	J_q^{\text{bare},(2)}(10)&=Z_{\alpha}^2 a_s^2 \left(\frac{4\mu^2}{Q^2 R^2}\right)^{2\epsilon}\Bigg[C_F T_F n_f\left(\frac{3.39662\times 10^{-7}}{\epsilon^3}+\frac{5.80395}{\epsilon^2}+\frac{62.9088}{\epsilon}\right)\notag\\
	&+C_F^2\left(-\frac{1.65019\times 10^{-7}}{\epsilon^4}+\frac{0.000239529}{\epsilon^3}+\frac{40.1611}{\epsilon^2}+\frac{372.38}{\epsilon}\right)\notag\\
	&+C_F C_A\left(\frac{9.59114\times 10^{-8}}{\epsilon^4}-\frac{0.000156933}{\epsilon^3}-\frac{17.1019}{\epsilon^2}-\frac{177.226}{\epsilon}\right)\Bigg]
\end{align}

\begin{figure}[!htbp]
	\includegraphics[scale=0.6]{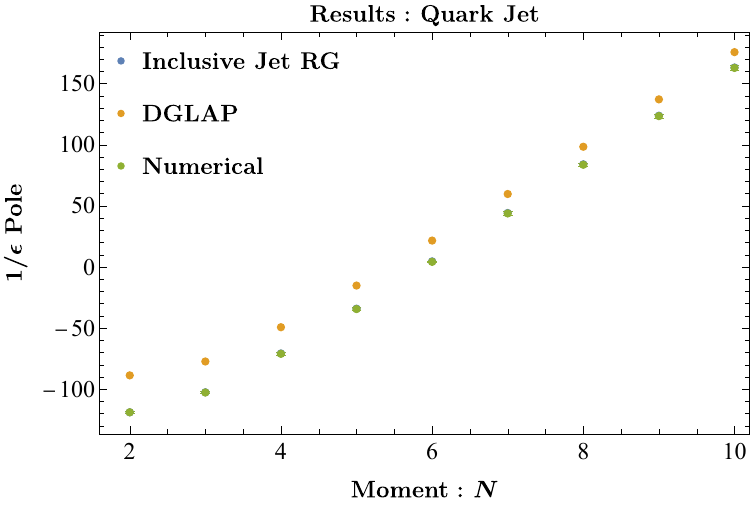}
	\includegraphics[scale=0.6]{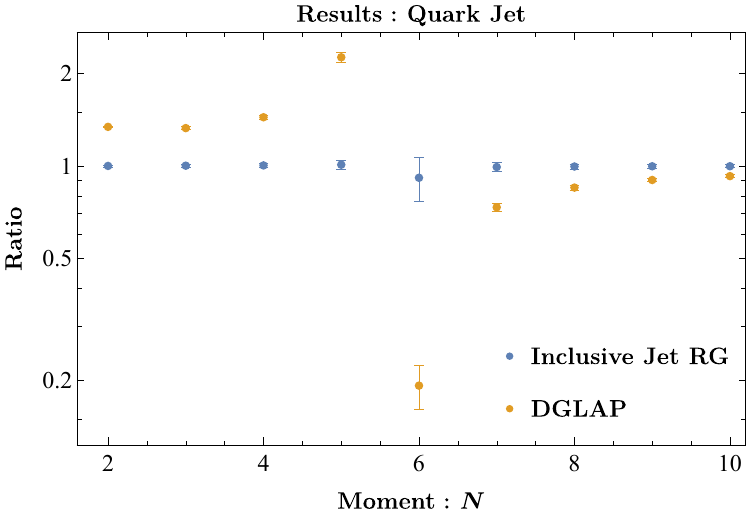}\\
 	\includegraphics[scale=0.6]{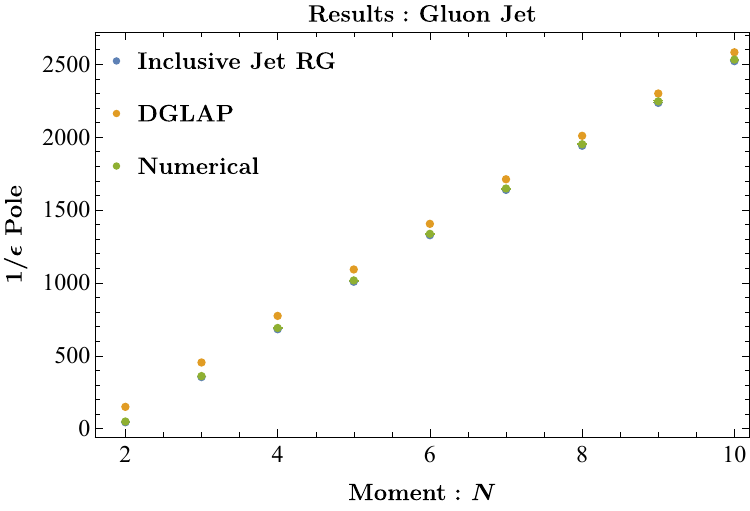}
	\includegraphics[scale=0.6]{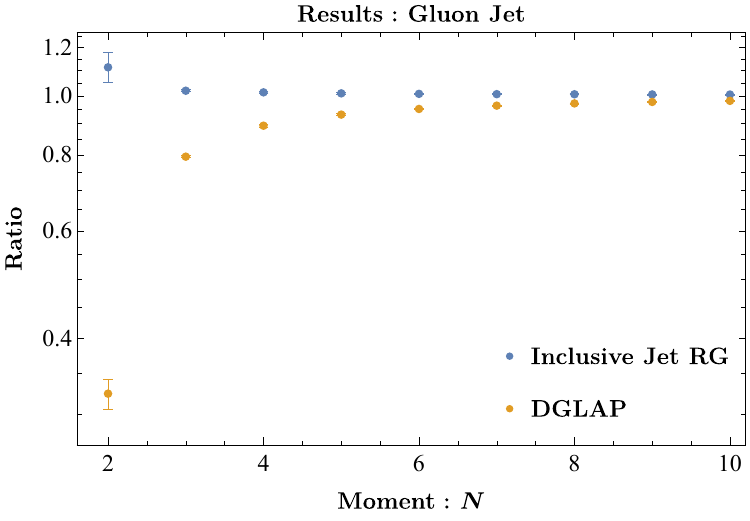}
	\caption{We compare the $1/\epsilon$ linear pole in the two-loop bare jet functions for QCD with the expected value derived from the inclusive jet RG and DGLAP, following a similar analysis to the $\mathcal{N}=4$ case shown in Fig.~\ref{fig:n4_plot}. For this comparison, we use $n_f = 5$, and find that our numerical results align much more closely with the predictions from the inclusive jet RG than with those from DGLAP. The linear poles for the $N=6$ quark jet and $N=2$ gluon jet are the smallest (closest to zero) among the moments considered, leading to larger relative errors in the ratio for these cases.}
	\label{fig:QCDjet_plot}
\end{figure}

\begin{figure}[!htbp]
	\includegraphics[scale=0.39]{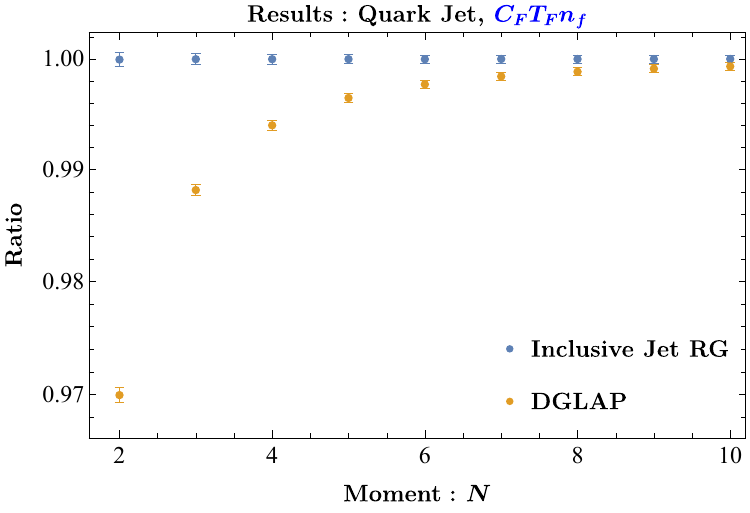}
	\includegraphics[scale=0.39]{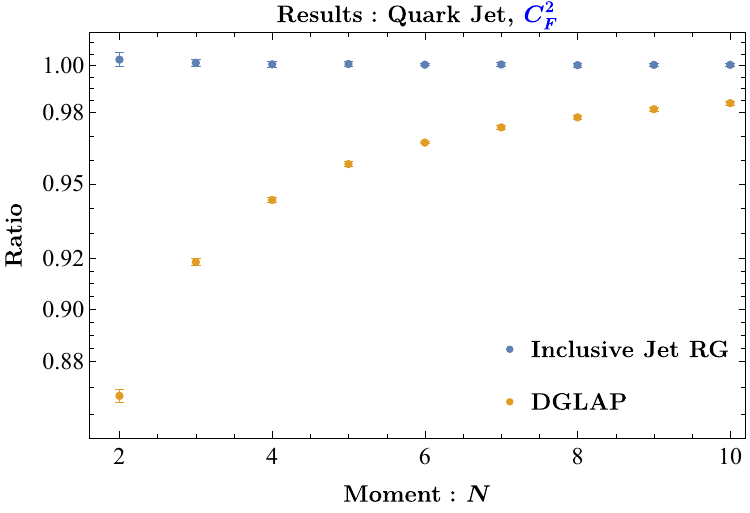}
	\includegraphics[scale=0.39]{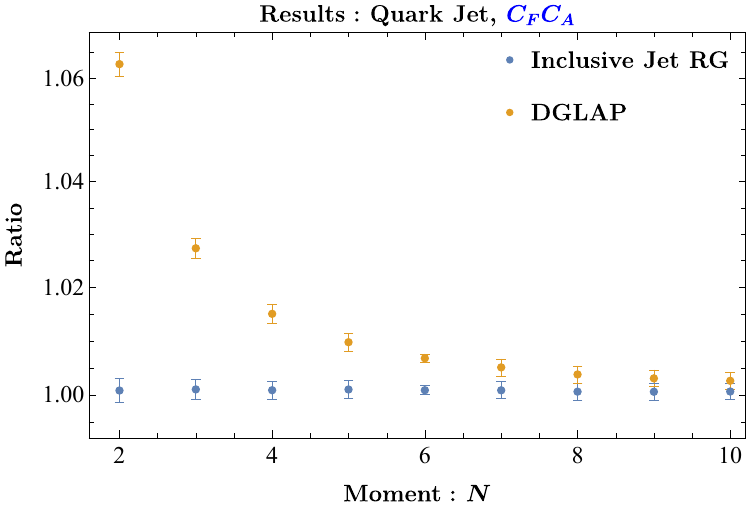}\\
 	\includegraphics[scale=0.39]{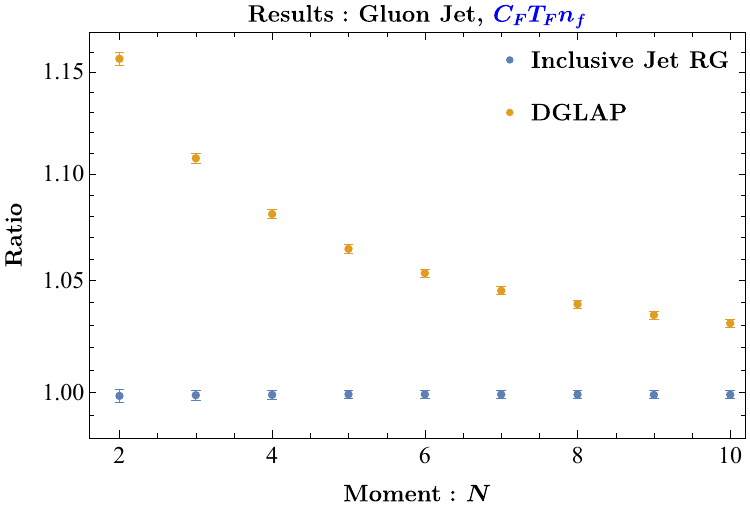}
	\includegraphics[scale=0.39]{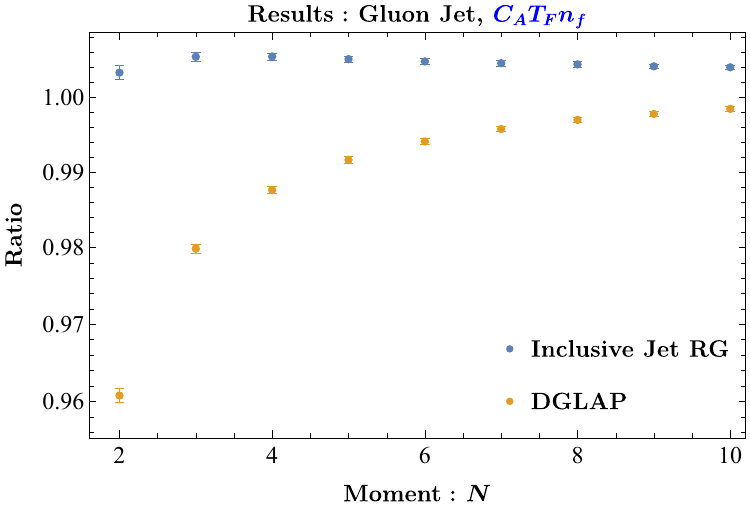}
	\includegraphics[scale=0.39]{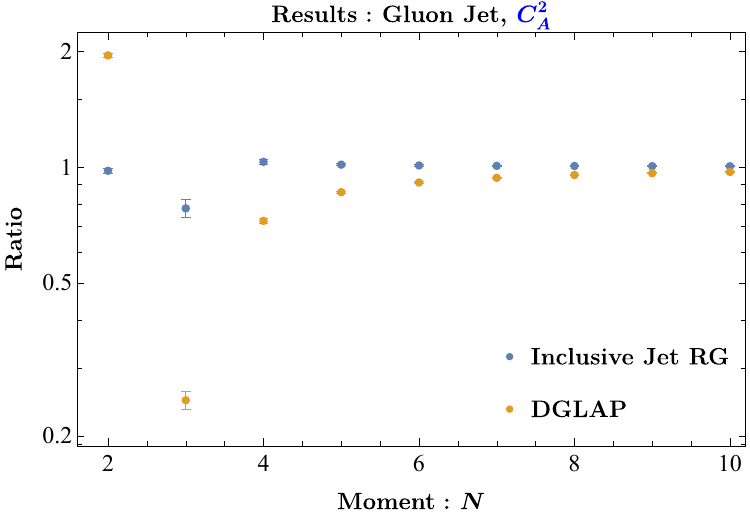}
	\caption{Ratio comparison of the $1/\epsilon$ linear pole across different color channels in the two-loop bare jet functions for QCD. In all channels, our numerical results show significantly closer alignment with the predictions from the inclusive jet RG compared to those from DGLAP.}
	\label{fig:QCDjet_plot_color}
\end{figure}

For the gluon jet, we have
\begin{align}
	J_g^{\text{bare},(2)}(1)&=Z_{\alpha}^2 a_s^2 \left(\frac{4\mu^2}{Q^2 R^2}\right)^{2\epsilon}\Bigg[C_F T_F n_f\left(\frac{3.78492\times 10^{-6}}{\epsilon^3}+\frac{0.00148535}{\epsilon^2}+\frac{0.0149253}{\epsilon}\right)\notag\\
	&+C_A T_F n_f\left(\frac{7.27394\times 10^{-7}}{\epsilon^3}+\frac{0.0134235}{\epsilon^2}-\frac{0.314216}{\epsilon}\right)\notag\\
	&+C_A^2\left(\frac{8.86927\times 10^{-9}}{\epsilon^4}+\frac{0.00146432}{\epsilon^3}+\frac{0.0401887}{\epsilon^2}+\frac{0.34237}{\epsilon}\right)\Bigg]\notag\\
	J_g^{\text{bare},(2)}(2)&=Z_{\alpha}^2 a_s^2 \left(\frac{4\mu^2}{Q^2 R^2}\right)^{2\epsilon}\Bigg[C_F T_F n_f\left(\frac{9.11259\times 10^{-7}}{\epsilon^3}-\frac{1.39865}{\epsilon^2}-\frac{11.6021}{\epsilon}\right)\notag\\
	&+C_A T_F n_f\left(-\frac{3.43247\times 10^{-6}}{\epsilon^3}+\frac{3.56453}{\epsilon^2}+\frac{32.2951}{\epsilon}\right)\notag\\
	&+C_A^2\left(\frac{8.83427\times 10^{-8}}{\epsilon^4}+\frac{0.00142669}{\epsilon^3}-\frac{1.18233}{\epsilon^2}-\frac{19.5442}{\epsilon}\right)\notag\\
	&+T_F^2 n_f^2 \left(\frac{0.533333}{\epsilon^2}+\frac{3.30667}{\epsilon}\right) \Bigg]\notag\\
	J_g^{\text{bare},(2)}(3)&=Z_{\alpha}^2 a_s^2 \left(\frac{4\mu^2}{Q^2 R^2}\right)^{2\epsilon}\Bigg[C_F T_F n_f\left(\frac{8.64777\times 10^{-7}}{\epsilon^3}-\frac{1.64901}{\epsilon^2}-\frac{13.9548}{\epsilon}\right)\notag\\
	&+C_A T_F n_f\left(-\frac{2.80201\times 10^{-6}}{\epsilon^3}+\frac{5.7614}{\epsilon^2}+\frac{52.2171}{\epsilon}\right)\notag\\
	&+C_A^2\left(\frac{1.28312\times 10^{-7}}{\epsilon^4}+\frac{0.00126039}{\epsilon^3}+\frac{1.16451}{\epsilon^2}-\frac{1.78162}{\epsilon}\right)\notag\\
	&+T_F^2 n_f^2 \left(\frac{0.8}{\epsilon^2}+\frac{4.96}{\epsilon}\right) \Bigg]\notag\\
	J_g^{\text{bare},(2)}(4)&=Z_{\alpha}^2 a_s^2 \left(\frac{4\mu^2}{Q^2 R^2}\right)^{2\epsilon}\Bigg[C_F T_F n_f\left(\frac{1.32878\times 10^{-6}}{\epsilon^3}-\frac{1.68545}{\epsilon^2}-\frac{14.7496}{\epsilon}\right)\notag\\
	&+C_A T_F n_f\left(-\frac{3.44182\times 10^{-6}}{\epsilon^3}+\frac{7.44022}{\epsilon^2}+\frac{68.0244}{\epsilon}\right)\notag\\
	&+C_A^2\left(\frac{1.68284\times 10^{-7}}{\epsilon^4}+\frac{0.00132871}{\epsilon^3}+\frac{3.93629}{\epsilon^2}+\frac{21.2511}{\epsilon}\right)\notag\\
	&+T_F^2 n_f^2 \left(\frac{0.965079}{\epsilon^2}+\frac{6.03187}{\epsilon}\right) \Bigg]\notag\\
	J_g^{\text{bare},(2)}(5)&=Z_{\alpha}^2 a_s^2 \left(\frac{4\mu^2}{Q^2 R^2}\right)^{2\epsilon}\Bigg[C_F T_F n_f\left(\frac{1.71009\times 10^{-6}}{\epsilon^3}-\frac{1.6579}{\epsilon^2}-\frac{15.0148}{\epsilon}\right)\notag\\
	&+C_A T_F n_f\left(-\frac{2.45099\times 10^{-6}}{\epsilon^3}+\frac{8.81988}{\epsilon^2}+\frac{81.5181}{\epsilon}\right)\notag\\
	&+C_A^2\left(\frac{2.08255\times 10^{-7}}{\epsilon^4}+\frac{0.00135695}{\epsilon^3}+\frac{6.74966}{\epsilon^2}+\frac{45.7953}{\epsilon}\right)\notag\\
	&+T_F^2 n_f^2 \left(\frac{1.07937}{\epsilon^2}+\frac{6.813}{\epsilon}\right) \Bigg]\notag\\
	J_g^{\text{bare},(2)}(6)&=Z_{\alpha}^2 a_s^2 \left(\frac{4\mu^2}{Q^2 R^2}\right)^{2\epsilon}\Bigg[C_F T_F n_f\left(\frac{1.59111\times 10^{-6}}{\epsilon^3}-\frac{1.60804}{\epsilon^2}-\frac{15.0373}{\epsilon}\right)\notag\\
	&+C_A T_F n_f\left(-\frac{2.34375\times 10^{-6}}{\epsilon^3}+\frac{9.99838}{\epsilon^2}+\frac{93.4421}{\epsilon}\right)\notag\\
	&+C_A^2\left(\frac{2.48235\times 10^{-7}}{\epsilon^4}+\frac{0.001516}{\epsilon^3}+\frac{9.5077}{\epsilon^2}+\frac{70.9369}{\epsilon}\right)\notag\\
	&+T_F^2 n_f^2 \left(\frac{1.16402}{\epsilon^2}+\frac{7.4196}{\epsilon}\right) \Bigg]\notag\\
	J_g^{\text{bare},(2)}(7)&=Z_{\alpha}^2 a_s^2 \left(\frac{4\mu^2}{Q^2 R^2}\right)^{2\epsilon}\Bigg[C_F T_F n_f\left(\frac{1.47067\times 10^{-6}}{\epsilon^3}-\frac{1.55105}{\epsilon^2}-\frac{14.9377}{\epsilon}\right)\notag\\
	&+C_A T_F n_f\left(-\frac{1.76463\times 10^{-6}}{\epsilon^3}+\frac{11.0294}{\epsilon^2}+\frac{104.191}{\epsilon}\right)\notag\\
	&+C_A^2\left(\frac{2.88241\times 10^{-7}}{\epsilon^4}+\frac{0.00137888}{\epsilon^3}+\frac{12.1741}{\epsilon^2}+\frac{96.1811}{\epsilon}\right)\notag\\
	&+T_F^2 n_f^2 \left(\frac{1.22963}{\epsilon^2}+\frac{7.90977}{\epsilon}\right) \Bigg]\notag\\
	J_g^{\text{bare},(2)}(8)&=Z_{\alpha}^2 a_s^2 \left(\frac{4\mu^2}{Q^2 R^2}\right)^{2\epsilon}\Bigg[C_F T_F n_f\left(\frac{1.41716\times 10^{-6}}{\epsilon^3}-\frac{1.4931}{\epsilon^2}-\frac{14.7728}{\epsilon}\right)\notag\\
	&+C_A T_F n_f\left(-\frac{2.67332\times 10^{-6}}{\epsilon^3}+\frac{11.9471}{\epsilon^2}+\frac{114.014}{\epsilon}\right)\notag\\
	&+C_A^2\left(\frac{3.28212\times 10^{-7}}{\epsilon^4}+\frac{0.00134625}{\epsilon^3}+\frac{14.7488}{\epsilon^2}+\frac{121.355}{\epsilon}\right)\notag\\
	&+T_F^2 n_f^2 \left(\frac{1.28215}{\epsilon^2}+\frac{8.31689}{\epsilon}\right) \Bigg]\notag\\
	J_g^{\text{bare},(2)}(9)&=Z_{\alpha}^2 a_s^2 \left(\frac{4\mu^2}{Q^2 R^2}\right)^{2\epsilon}\Bigg[C_F T_F n_f\left(\frac{1.46684\times 10^{-6}}{\epsilon^3}-\frac{1.43658}{\epsilon^2}-\frac{14.5707}{\epsilon}\right)\notag\\
	&+C_A T_F n_f\left(-\frac{2.73677\times 10^{-6}}{\epsilon^3}+\frac{12.7731}{\epsilon^2}+\frac{123.06}{\epsilon}\right)\notag\\
	&+C_A^2\left(\frac{3.6819\times 10^{-7}}{\epsilon^4}+\frac{0.00145695}{\epsilon^3}+\frac{17.2252}{\epsilon^2}+\frac{146.202}{\epsilon}\right)\notag\\
	&+T_F^2 n_f^2 \left(\frac{1.32525}{\epsilon^2}+\frac{8.66199}{\epsilon}\right) \Bigg]\,,\notag\\
	J_g^{\text{bare},(2)}(10)&=Z_{\alpha}^2 a_s^2 \left(\frac{4\mu^2}{Q^2 R^2}\right)^{2\epsilon}\Bigg[C_F T_F n_f\left(\frac{1.51709\times 10^{-6}}{\epsilon^3}-\frac{1.38296}{\epsilon^2}-\frac{14.3534}{\epsilon}\right)\notag\\
	&+C_A T_F n_f\left(-\frac{3.00455\times 10^{-6}}{\epsilon^3}+\frac{13.5255}{\epsilon^2}+\frac{131.479}{\epsilon}\right)\notag\\
	&+C_A^2\left(\frac{4.08161\times 10^{-7}}{\epsilon^4}+\frac{0.00156466}{\epsilon^3}+\frac{19.6091}{\epsilon^2}+\frac{170.725}{\epsilon}\right)\notag\\
	&+T_F^2 n_f^2 \left(\frac{1.36131}{\epsilon^2}+\frac{8.95918}{\epsilon}\right) \Bigg]\,.
\end{align}

Similar to the $\mathcal{N}=4$ case, we find that all quartic and cubic poles cancel within small and similarly sized numerical effects across all moments when the virtual and real corrections are combined. The $N=1$ moment vanishes identically in the QCD case according to the RG prediction, and the Monte Carlo results also confirm this up to small numerical effects. For $N \geq 2$ moments, the quadratic pole agrees with our RG prediction from~\eq{Jpred}, with a similarly small numerical deviation. As noted earlier for the $\mathcal{N}=4$ case, both the inclusive jet RG and DGLAP yield the same predictions for the quadratic pole. Thus, to differentiate between the two, we must analyze the linear pole.

In Fig.~\ref{fig:QCDjet_plot} and~\ref{fig:QCDjet_plot_color}, we compare the numerically computed $1/\epsilon$ linear pole with what is expected from the inclusive jet RG and DGLAP evolutions, for both quark and gluon jets. We find that across all moments and all color channels, our numerical results are consistent with what is expected from the inclusive jet RG with modified convolution structure. In the forthcoming paper, we will also explore the numerical impact of the differences between the inclusive jet RG and DGLAP evolutions, which already become important at NLL resummation accuracy.

\subsection{Exclusive Jet Production}
\label{sec:excljet}
For our single-inclusive jet production, it is important to allow the production of multiple jets in the final state, though the transverse momentum or energy fraction of the jet is measured only once at a time. This necessitates the computation of mult-jet final states as shown in Figs.~\ref{fig:virtual_config} and~\ref{fig:ft}. In contrast, exclusive jet production strictly prohibits multi-jet final states. Exclusive jet production process is phenomenologically important, including for separating different S-matrix elements of the Higgs processes by $H+n$ jets~\cite{Cal:2024yjz,Liu:2013hba}. This exclusive jet restriction can be imposed through jet veto condition, which is associated with soft-collinear modes in the exclusive jet factorization~\cite{Cal:2024yjz,Chien:2015cka}. Additionally, the exclusive jet factorization contains an associated jet function, which we refer to as the ``exclusive jet function'' here, describing the situation where all final state particles are clustered into a single jet. Such an exclusive jet function was computed for the quark case at two loops for the first time in~\cite{Liu:2021xzi}. Our calculation naturally includes the exclusive jet function as a subset, and we compare our results to the quark case discussed in~\cite{Liu:2021xzi} and further report the result for the gluon. 

The exclusive jet function follows the renormalization group evolution equation~\cite{Ellis:2010rwa}
\begin{align}
    \frac{d J_i^{\rm{ex}}}{d\ln \mu^2}=\frac{1}{2}\left(\Gamma_i^{\rm cusp} L+\gamma_{i}\right)J_i^{\rm ex}\,,
\end{align}
where $L = \ln [4\mu^2/(Q^2R^2)]$\,. The $\Gamma_i^{\rm cusp}$ and $\gamma_i$ denote cusp and non-cusp anomalous dimensions, respectively, which can be expanded in $a_s = \alpha_s/(4\pi)$ as 
\begin{align}
\Gamma^{\rm{cusp}}_i&=\sum_{n=1}^{\infty} a_s^n \Gamma^{n-1}_i\,,\qquad
\gamma_i =\sum_{n=1}^{\infty} a_s^n \gamma^{n-1}_i\,.
\end{align}
The cusp anomalous dimensions are known to four loops~\cite{Korchemsky:1987wg,Moch:2004pa,Henn:2019swt,vonManteuffel:2020vjv}. To two loops, they are given as
\begin{align}
\Gamma^0_i=4C_i 
      \,, \quad& \Gamma^1_i=4C_i\left[C_A\left(\frac{67}{9}-\frac{\pi^2}{3}\right)-\frac{20}{9}T_F n_F\right]\,.
\end{align}
Currently, only one loop non-cusp anomalous dimension is fully known for the exclusive jet function is given as~\cite{Ellis:2010rwa}
\begin{align}
\gamma_q^0 = 6 C_F\,, \quad \gamma_g^0 = 2\beta_0\,.
\end{align}
The RG predicts the renormalized exclusive jet function to take the form
\begin{align}\label{eq:excl_renor_exp}
    J_i^{\rm{ex}}&=1+a_s\Bigg[\frac{1}{4}\Gamma_i^0 L^2+\frac{1}{2}\gamma_i^0+J_{i,0}^{\rm{ex}(1)}L\Bigg]+a_s^2\Bigg[\frac{(\Gamma_i^0)^2}{32}L^4+\frac{\Gamma_i^0}{24}\left(2\beta_0+3\gamma_i^0\right)L^3\nn\\
    &+\frac{1}{8}\left(2\beta_0\gamma_i^0+(\gamma_i^0)^2+2J_{i,0}^{\rm{ex}(1)}\Gamma_i^0+2\Gamma_i^1\right)L^2+\frac{1}{2}\left(2J_{i,0}^{\rm{ex}(1)}\beta_0+J_{i,0}^{\rm{ex}(1)}\gamma_i^0+\gamma_i^1\right)L+J_{i,0}^{\rm{ex}(2)}\Bigg]\nn\\
    &+\mathcal{O}(a_s^3)
\end{align}
The two-loop non-cusp anomalous dimension was computed for the first time in~\cite{Liu:2021xzi} for exclusive quark jet, which we will comment on below.

On the other hand, the bare exclusive jet function can be expressed as
\begin{align}\label{eq:excl_bare_exp}
J_{i,\rm{bare}}^{\rm{ex}}&=1+Z_{\alpha}a_s \left(\frac{4\mu^2}{Q^2 R^2}\right)^{\epsilon}\left(\frac{J_{i,-2}^{\rm{ex}(1)}}{\epsilon^2}+\frac{J_{i,-1}^{\rm{ex}(1)}}{\epsilon}+J_{i,0}^{\rm{ex}(1)}+J_{i,1}^{\rm{ex}(1)}\epsilon\right)\nn\\
&+Z_{\alpha}^2 a_s^2 \left(\frac{4\mu^2}{Q^2 R^2}\right)^{2\epsilon}\left(\frac{J_{i,-4}^{\rm{ex}(2)}}{\epsilon^4}+\frac{J_{i,-3}^{\rm{ex}(2)}}{\epsilon^3}+\frac{J_{i,-2}^{\rm{ex}(2)}}{\epsilon^2}+\frac{J_{i,-1}^{\rm{ex}(2)}}{\epsilon}\right)\,.
\end{align}
Renormalizing this bare jet function then gives the renormalized expression as
\begin{align}
\label{eq:renExcl2}
J_{i}^{\rm{ex}}=1&+a_s \Bigg[\frac{1}{2}J_{i,-2}^{\rm{ex}(1)}L^2+J_{i,-1}^{\rm{ex}(1)}L+J_{i,0}^{\rm{ex}(1)}\Bigg]\nn\\
&+a_s^2\Bigg[\left(\frac{2}{3}J_{i,-4}^{\rm{ex}(2)}-\frac{5}{24}(J_{i,-2}^{\rm{ex}(1)})^2\right)L^4+\left(\frac{4}{3}J_{i,-3}^{\rm{ex}(2)}-\frac{\beta_0}{6} J_{i,-2}^{\rm{ex}(1)}-\frac{5}{6}J_{i,-2}^{\rm{ex}(1)}J_{i,-1}^{\rm{ex}(1)}\right)L^3\nn\\
&+\left(2J_{i,-2}^{\rm{ex}(2)}-\frac{\beta_0}{2}J_{i,-1}^{\rm{ex}(1)}-\frac{1}{2}(J_{i,-1}^{\rm{ex}(1)})^2-\frac{3}{2}J_{i,-2}^{\rm{ex}(1)}J_{i,0}^{\rm{ex}(1)}\right)L^2\nn\\
&+\left(2J_{i,-1}^{\rm{ex}(2)}-\beta_0 J_{i,0}^{\rm{ex}(1)}-J_{i,-1}^{\rm{ex}(1)}J_{i,0}^{\rm{ex}(1)}-2J_{i,-2}^{\rm{ex}(1)}J_{i,1}^{\rm{ex}(1)}\right)L+J_{i,0}^{\rm{ex}(2)}\Bigg] \,.
\end{align}
Comparing~\eq{excl_renor_exp} with~\eq{renExcl2}, we are able to determine the associated divergences as
\begin{align}
\label{eq:RGexpExcl}
    J_{i,-2}^{\rm{ex}(1)}&=\frac{1}{2}\Gamma_i^0\,, &J_{i,-1}^{\rm{ex}(1)}&=\frac{1}{2}\gamma_i^0\,,\\
    J_{i,-4}^{\rm{ex}(2)}&=\frac{1}{8}(\Gamma_i^0)^2\,, &J_{i,-3}^{\rm{ex}(2)}&=\frac{1}{8}\Gamma_i^0\beta_0+\frac{1}{4}\gamma_i^0\,,\nn\\
    J_{i,-2}^{\rm{ex}(2)}&=\frac{1}{8}\Gamma_i^1+\frac{1}{2}\Gamma_i^0 J_{i,0}^{\rm{ex}(1)}+\frac{1}{8}(\gamma_i^0)^2+\frac{1}{4}\gamma_i^0 \beta_0\,, &J_{i,-1}^{\rm{ex}(2)} &=\beta_0 J_{i,0}^{\rm{ex}(1)}+\frac{1}{4}\gamma_i^{1}+\frac{1}{2}\gamma_i^{0}J_{i,0}^{\rm{ex}(1)}+\frac{1}{2}\Gamma_i^{0}J_{i,1}^{\rm{ex}(1)}\nn
\end{align}

Now we are ready to compare with our calculation. For both one loop exclusive jet function and for its virtual corrections, we have two final state particles. For exclusive jet function measurement, we only need the case where both particles are clustered together, which corresponds to the first term in Eq.~\eqref{eq:meas2} and Eq.~\eqref{eq:Jvirtmoments}. For real corrections, we only need the configuration where all three particles are inside the same jet. This corresponds to the case (A) measurement discussed in~\eq{measA}. Combining these calculations gives the bare exclusive jet function to two loops for quark jet
\begin{align}\label{eq:qjet_excl}
    J_{q,{\rm bare}}^{\rm{ex}}&=Z_{\alpha}a_s \left(\frac{4\mu^2}{Q^2 R^2}\right)^\epsilon C_F\Bigg[\frac{2}{\epsilon^2}+\frac{3}{\epsilon}+13-9\zeta_2+\left(52-\frac{27}{2}\zeta_2-\frac{98}{3}\zeta_3\right)\epsilon\Bigg]\notag\\
    &+Z_{\alpha}^2 a_s^2 \left(\frac{4\mu^2}{Q^2 R^2}\right)^{2\epsilon}\Bigg[C_F T_F n_f\left(-\frac{0.666666}{\epsilon^3}-\frac{3.11085}{\epsilon^2}+\frac{0.426687}{\epsilon}\right)\notag\\
	&+C_F^2\left(\frac{2}{\epsilon^4}+\frac{5.99928}{\epsilon^3}+\frac{0.889727}{\epsilon^2}-\frac{{\color{red}16.9332}}{\epsilon}\right)\notag\\
	&+C_F C_A\left(-\frac{3.49603\times 10^{-8}}{\epsilon^4}+\frac{1.83289}{\epsilon^3}+\frac{4.2792}{\epsilon^2}-\frac{52.3257}{\epsilon}\right)\Bigg]+\mathcal{O}(\alpha_s^3)\,,
\end{align}
and also for gluon jet
\begin{align}\label{eq:gjet_excl}
J_{g,{\rm bare}}^{\rm{ex}}&=Z_{\alpha} a_s \left(\frac{4\mu^2}{Q^2 R^2}\right)^{\epsilon}\Bigg[T_F n_f \left(-\frac{4}{3\epsilon}-\frac{46}{9}+\epsilon\left(-\frac{562}{27}+6\zeta_2\right)\right)\notag\\
&+C_A\left(\frac{4}{\epsilon^2}+\frac{22}{3\epsilon}+\frac{268}{9}-18\zeta_2+\epsilon\left(\frac{3232}{27}-33\zeta_2-\frac{196}{3}\zeta_3\right)\right)\Bigg]\notag\\
&+Z_{\alpha}^2 a_s^2 \left(\frac{4\mu^2}{Q^2 R^2}\right)^{2\epsilon}\Bigg[C_F T_F n_f\left(\frac{3.53707\times 10^{-6}}{\epsilon^3}+\frac{0.00125938}{\epsilon^2}+\frac{1.12679}{\epsilon}\right)\notag\\
	&+T_F^2 n_f^2 \left(\frac{16}{9\epsilon^2}+\frac{368}{27\epsilon}\right)+C_A T_F n_f\left(-\frac{3.33333}{\epsilon^3}-\frac{21.1105}{\epsilon^2}-\frac{59.312}{\epsilon}\right)\notag\\
	&+C_A^2\left(\frac{2}{\epsilon^4}+\frac{9.16555}{\epsilon^3}+\frac{12.3903}{\epsilon^2}-\frac{51.1444}{\epsilon}\right) \Bigg]+\mathcal{O}(\alpha_s^3)\,.
\end{align}
The two-loop constant part of the exclusive jet function will be presented in a forthcoming paper. These terms are often a source of leading perturbative uncertainty, as seen in the recent $H+1$ jet calculations in \cite{Cal:2024yjz}. The two-loop exclusive quark jet function was computed in~\cite{Liu:2021xzi} as noted above. Our results are in close agreement with the result of~\cite{Liu:2021xzi}, except for the single pole in $C_F^2$ indicated in red in~\eq{qjet_excl}. Our result is given as $-16.9332$, compared to the value given in~\cite{Liu:2021xzi}, which is $-21.5692$. Due to many consistency checks with our inclusive jet RG, which takes our exclusive jet function as a subset, we are confident in our value. On the other hand, our two-loop exclusive gluon jet function is a completely new calculation. 

Compared to what is expected from the exclusive jet function RG given in~\eq{RGexpExcl}, our computation of the coefficients $J_{i,-4}^{(2)}$ and $J_{i,-3}^{(2)}$ are in very close agreement. For $1/\epsilon^2$, we find good agreement except for $C_F C_A$ channel in the quark jet and $C_A^2$ in the gluon jet. These terms receive additional non-global contributions for exclusive jet production~\cite{Liu:2021xzi,Dasgupta:2001sh,Becher:2016mmh} which modifies
\begin{align}
J_{i,-2}^{(2)}|_{C_i C_A} \to J_{i,-2}^{(2)}|_{C_i C_A} - C_i C_A \frac{\pi^2}{3}\,.
\end{align}
Accounting for NGL contributions, our results for the double pole is also in agreement.

Finally, the value of the single pole is related to the two-loop non-cusp anomalous dimension $\gamma_i^1$ as shown in~\eq{RGexpExcl}. This relation will hold up to contribution from subleading NGL, which we ignore for now. This allows us to derive the two-loop non-cusp anomalous dimension $\gamma_i^1$ for both quark and gluon exclusive jet function as
\begin{align}
\gamma_q^{(1)} =& 29.7105 C_F^2-182.838 C_F C_A-7.91675C_F T_F n_f \,,\nn\\
\gamma_g^{(1)} =&-104.626 C_A^2+4.50716 C_F T_F n_f +2.0419C_A T_F n_f \,.
\end{align}
The two-loop non-cusp anomalous dimension for the gluon jet is a novel result, while for the quark jet, it is different for the $C_F^2$ color channel compared to the value $11.17(5) C_F^2$ from~\cite{Liu:2021xzi}. Together with the two-loop constant terms of the jet function, which will be detailed in our forthcoming paper, these results establish a foundation for precision in both inclusive and exclusive jet processes.

\section{Conclusions}\label{sec:conc}

The inclusive production of jets is one of the most fundamental processes in all collider experiments and is used at the LHC in both proton-proton and heavy ion collisions, in DIS experiments for precision studies of nuclear structure, and in $e^+e^-$ colliders for precision jet measurements. Essential to accurate predictions of inclusive jet production is their factorization theorems, which also become important for the analysis of jet substructure observables. While factorization theorems for spacelike processes like DIS are well-established, their timelike counterparts—beyond the simplest case of inclusive hadron fragmentation—remain less explored at higher perturbative orders.

In this paper, we build on recent advances in the factorization of single-logarithmic observables, particularly that of projected energy correlators, to present a novel factorization theorem for timelike process of inclusive jet production. Our factorization theorem introduces a non-trivial convolution structure compared to standard timelike inclusive hadron fragmentation but remains entirely determined by the timelike DGLAP anomalous dimensions and their derivatives. Notably, the moments of the inclusive jet process are shown to satisfy the exact same factorization theorem as the projected energy correlators. Our new factorization theorems highlight how explicit perturbative calculations and a deeper understanding of formal properties of anomalous dimensions in field theory can significantly impact the collider program.

To test our factorization, we have performed an explicit two-loop calculation of the anomalous dimension of the inclusive jet functions in both $\mathcal{N}=4$ super Yang-Mills theory and for all color channels in QCD. We find excellent agreement with the results of our RG analysis, validating our factorization theorem. This calculation also reveals the inconsistency of the previously proposed factorization formula stated in \cite{Kang:2016mcy,Kang:2016ehg}. While our results for the two-loop calculation agree with those in \cite{vanBeekveld:2024jnx}, for the channels they considered, our interpretation differs. Crucially, the IR measurement can modify the convolution structure in the factorization theorem but should not alter the anomalous dimensions, which are UV in origin. Our explicit calculations confirm this principle. A key advantage of our factorization theorem is that it preserves the universality of the hard function and its DGLAP evolution, allowing us to leverage known results on DGLAP anomalous dimensions to achieve higher logarithmic accuracy.

We further extend our factorization to various jet substructure measurements in the single-inclusive jet production process and utilize the close connection between moments of the inclusive jet process and projected energy correlators to develop a new jet algorithm that serves as a generating function for the projected energy correlators. Our results also provide the first two-loop calculation of the anomalous dimensions for the gluon exclusive jet function and independently confirm and slightly modify the result for the two-loop quark exclusive jet function from~\cite{Liu:2021xzi}.

Our results lay the groundwork for achieving greater precision in various jet production and substructure analyses, with the most immediate application in inclusive jet production itself. NLL predictions using the formalism of \cite{Kang:2016mcy,Kang:2016ehg} have been used, for example, in \cite{Chen:2021uws}, but these must be corrected in light of our findings, and it will be interesting to assess the numerical size of the effects. Beyond this, the most pressing direction is the extension to NNLL. This requires the calculation of the two-loop constants for the inclusive jet functions. For this purpose, it will also be interesting to explore the formal properties of the inclusive jet functions, for example, in the threshold or small-x regions. This will allow us to combine two-loop results for the energy correlator jet functions to provide NNLL single-logarithmic precision in jet substructure at hadron colliders for the first time and significantly improve the perturbative accuracy of the jet substructure program.

\acknowledgments

We thank Xiaohui Liu for sharing the calculation code in Ref.~\cite{Liu:2021xzi} and for many helpful discussions. We thank Zhongbo Kang, Yibei Li, Matthew Lim, and Wouter Waalewijn for helpful discussions. K.L. was supported by the U.S. Department of Energy, Office of Science, Office of Nuclear Physics from DE-SC0011090. I.M. is supported by the DOE Early Career Award DE-SC0025581. XY.Z. was supported in part by the U.S. Department of Energy under contract DE-SC0013607.

\appendix

\section{Perturbative Ingredients}\label{sec:pert}

In this Appendix we summarize the perturbative ingredients required for the two-loop calculation of the inclusive jet functions.

The one-loop corrections to the $1\to 2$ splittings functions, computed in conventional dimensional regularization, are \cite{Kosower:1999rx,Sborlini:2013jba}\footnote{Note that as compared to \cite{Sborlini:2013jba}, we have corrected a typo identified in \cite{Chen:2022pdu}. In the $g\to q\bar q$ splitting, the denominator is $4(\epsilon-2)\epsilon+3$, instead of $4(\epsilon-2)\epsilon-3$.}
\begin{align}
    P^{(1)}_{q\to qg}(x;s)
    &=
    2a_s\Bigl(\frac{\mu^2e^{\gamma_E}}{s}\Bigr)^\epsilon
    \frac{\pi\ \Gamma(1-\epsilon)}{\epsilon\tan{(\pi\epsilon)}\Gamma(1-2\epsilon)}\
    \biggl\{P^{(0)}_{q\to qg}(x)\biggl[C_F
    +(C_F-C_A)\Bigl(1-\frac{\epsilon^2}{1-2\epsilon}\Bigr)\nn
    \\
    &\qquad\qquad
    +(C_A-2C_F)\,{_2}F_1\Bigl(1,-\epsilon;1-\epsilon;\frac{x-1}{x}\Bigr)
    -C_A\,{_2}F_1\Bigl(1,-\epsilon;1-\epsilon;\frac{x}{x-1}\Bigr)\biggr]\nn
    \\
    &\qquad+C_F(C_F-C_A)\frac{x(1+x)}{1-x}\frac{\epsilon^2}{1-2\epsilon}
    \biggr\}
    \\[2ex]
    P^{(1)}_{g\to q \bar{q}}(x;s)
    &=
    2a_s\Bigl(\frac{\mu^2e^{\gamma_E}}{s}\Bigr)^\epsilon
    \frac{\pi\ \Gamma(1-\epsilon)}{\epsilon\tan{(\pi\epsilon)}\Gamma(1-2\epsilon)}\
    P^{(0)}_{g\to q\bar{q}}(x)
    \biggl\{
    C_A \biggl[\frac{[2 (\epsilon -2) \epsilon +2 \epsilon -3]
    \epsilon ^2+3}{(3-2 \epsilon )(\epsilon -1)(2\epsilon-1)}\nn
    \\
    &\qquad\qquad+2
    -\,{_2}F_1\Bigl(1,-\epsilon;1-\epsilon;\frac{x-1}{x}\Bigr)
    -\,{_2}F_1\Bigl(1,-\epsilon;1-\epsilon;\frac{x}{x-1}\Bigr)\biggr]
    \\
    &\qquad+C_F\,\frac{\epsilon  \left(2 \epsilon ^2-3 \epsilon +3\right)-2}{(\epsilon -1) (2 \epsilon -1)}
    +n_f T_F\,\frac{4 (\epsilon -1) \epsilon}{4
    (\epsilon -2) \epsilon +3}\biggr\}\ ,\nn
    \\[2ex]
    P^{(1)}_{g\to gg}(x;s)
    &=
    2a_s\Bigl(\frac{\mu^2e^{\gamma_E}}{s}\Bigr)^\epsilon
    \frac{\pi\ \Gamma(1-\epsilon)}{\epsilon\tan{(\pi\epsilon)}\Gamma(1-2\epsilon)}\
    C_A \biggl\{\frac{\epsilon^2 [1-2x(1-x)\epsilon] [C_A(\epsilon-1)+2n_fT_F]}
        {(1-\epsilon)(\epsilon-1)(2\epsilon-3) (2\epsilon-1)}\nn
    \\
    &\qquad+P^{(0)}_{g\to gg}(x)\biggl[1
    -\,{_2}F_1\Bigl(1,-\epsilon;1-\epsilon;\frac{x-1}{x}\Bigr)
    -\,{_2}F_1\Bigl(1,-\epsilon;1-\epsilon;\frac{x}{x-1}\Bigr)\biggr]\biggr\}\ .
\end{align}

We also present the complete set of $1\to 3$ splitting functions \cite{Campbell:1997hg,Catani:1998nv}
\begin{align}
    &P^{(0)}_{q\to qQ\bar{Q}}(x_i,x_j,x_k;s_{ijk},s_{ij},s_{ik},s_{jk})
    \\
    &=C_F T_F \frac{s_{ijk}}{2s_{jk}}\biggl\{-\frac{1}{s_{ijk}\ s_{jk}}
    \biggl[\frac{s_{jk}(x_j-x_k)}{x_j+x_k}
    +\frac{2(s_{ik}x_j-s_{ij}x_k)}{x_j+x_k}\biggr]^2\nn
    +(1-2\epsilon)\Bigl(-\frac{s_{jk}}{s}+x_j+x_k\Bigr)\nn
    \\
    &\qquad\quad
    +\frac{(x_j-x_k)^2+4x_i}{x_j+x_k}\biggr\}\ ,\nn
    \\[2ex]
    &P^{(0)}_{q\to qq\bar{q}}(x_i,x_j,x_k;s_{ijk},s_{ij},s_{ik},s_{jk})
    \\
    &=C_F\bigl(C_F-\tfrac{1}{2}C_A\bigr)\biggl\{
    -\frac{s_{ijk}^2}{2s_{jk}s_{ik}} x_k\biggl[\frac{1+x_k^2}{(1-x_j)(1-x_i)}
    -\epsilon\biggl(\frac{2(1-x_j)}{1-x_i}+1\biggr)-\epsilon^2\biggr]\nn
    \\
    &\qquad\quad
    +\frac{s_{ijk}}{s_{jk}}\biggl[\frac{1+x_k^2}{1-x_j}
    -\epsilon\biggl(1+x_k+\frac{(1-x_i)^2}{1-x_j}-\frac{2x_j}{1-x_i}\biggr)
    -\frac{2x_j}{1-x_i}-\epsilon^2(1-x_i)\biggr]\nn
    \\
    &\qquad\quad
    +(1-\epsilon)\Bigl(\frac{2s_{ij}}{s_{jk}}-\epsilon\Bigr)\biggr\}
    +P^{(0)}_{q\to qQ\bar{Q}}(x_i,x_j,x_k;s_{ijk},s_{ij},s_{ik},s_{jk})
    +(1\leftrightarrow2)\ ,\nn
    \\[2ex]
    &P^{(0)}_{q\to qgg}(x_i,x_j,x_k;s_{ijk},s_{ij},s_{ik},s_{jk})
    \\
    &=C_F^2\biggl\{
    \frac{s_{ijk}^2}{2s_{ik}s_{ij}} x_i\biggl[\frac{1+x_i^2}{x_j x_k}
    -\epsilon\frac{x_j^2+x_k^2}{x_j x_k}-\epsilon(1+\epsilon)\biggr]
    +(1-\epsilon)\Bigl(\epsilon-(1-\epsilon)\frac{s_{ij}}{s_{ik}}\Bigr)\nn
    \\
    &\qquad\quad
    +\frac{s_{ijk}}{s_{ik}}\biggl[\frac{(1-x_k)x_i+(1-x_j)^3}{x_j x_k}
    -\epsilon\frac{(1-x_j)(x_j^2+x_j x_k+x_k^2)}{x_j x_k}
    +(1+x_i)\epsilon^2\biggr]
    \biggr\}\nn
    \\
    &\quad
    +C_A C_F\biggl\{\frac{s_{ijk}^2}{2s_{jk} s_{ik}}
    \biggl[\frac{x_j^2 (1-\epsilon)+2(1-x_j)}{1-x_i}
    +\frac{2x_i+(1-\epsilon)(1-x_i)^2}{x_j}\biggr]\nn
    \\
    &\qquad\quad
    -\frac{s_{ijk}^2}{4s_{ij} s_{ik}} x_i 
    \biggl[\frac{(1-x_i)^2(1-\epsilon)+2x_i}{x_j x_k}
    +\epsilon(1-\epsilon)\biggr]\nn
    \\
    &\qquad\quad
    +\frac{s_{ijk}}{2s_{jk}}\biggl[(1-\epsilon)
    \frac{x_k(x_k^2-2 x_k+2)-x_j(x_j^2-6 x_j+6)}{x_j(1-x_i)}
    +2\epsilon\frac{x_i(x_k-2 x_j) - x_j}{x_j (1-x_i)}\biggr]\nn
    \\
    &\qquad\quad
    +\frac{s_{ijk}}{2s_{ik}}\biggl[\epsilon(1-x_j)
    \Bigl(\frac{x_j^2+x_k^2}{x_j x_k}-\epsilon\Bigr)
    -\epsilon\Bigl(x_j-x_k+\frac{2(1-x_j)(x_j-x_i)}{x_j (1-x_i)}\Bigr)\nn
    \\
    &\qquad\qquad\quad
    -\frac{(1-x_k)x_i+(1-x_j)^3}{x_j x_k}
    +(1-\epsilon)\frac{(1-x_j)^3-x_j+x_i^2}{x_j(1-x_i)}\biggr]\nn
    \\
    &\qquad\quad
    +(1-\epsilon)\biggl[\frac{1}{4s_{jk}^2}\biggl(\frac{s_{jk}(x_k-x_j)}{x_j+x_k}
    +\frac{2(s_{ij}x_k-s_{ik} x_j)}{x_j+x_k}\biggr)^2
    -\frac{\epsilon}{2}+\frac{1}{4}\biggr]\biggr\}+(2\leftrightarrow3)\nn
    \\[2ex]
    &P^{(0)}_{g\to gq\bar{q}}(x_i,x_j,x_k;s_{ijk},s_{ij},s_{ik},s_{jk})
    \\
    &=C_F T_F\biggl\{
    \frac{s_{ijk}^2}{s_{ij} s_{ik}}\Bigl(x_i^2-\frac{x_i+2x_j x_k}{1-\epsilon}+1\Bigr)
    -\frac{s_{ijk}}{s_{ij}}\Bigl(2x_i-\frac{2(x_i+x_j)}{1-\epsilon}+1+\epsilon\Bigr)
    -(1-\epsilon)\frac{s_{jk}}{s_{ij}}-1\biggr\}\nn
    \\
    &\quad+
    C_A T_F\biggl\{-\frac{s_{ijk}^2}{2 s_{ij} s_{ik}}
    \biggl[1+x_i^2-\frac{x_i+2 x_j x_k}{1-\epsilon}\biggr]
    +\frac{s_{ijk}^2}{2s_{ik} s_{jk}}x_k\biggl[\frac{(1-x_i)^3-x_i^3}{(1-x_i) x_i}
    -\frac{2x_k (1-2x_i x_j-x_k)}{(1-x_i) x_i (1-\epsilon )}\biggr]\nn
    \\
    &\qquad\quad
    +\frac{s_{ijk}}{2s_{ik}}(1\!-\!x_j)\biggl[\frac{1+x_i-x_i^2}{(1\!-\!x_i) x_i}
    -\frac{2 (1\!-\!x_j)x_j}{(1\!-\!x_i) x_i (1\!-\!\epsilon)}\biggr]
    -\frac{1}{4s_{jk}^2}\biggl[\frac{2 (s_{ik}x_j-s_{ij} x_k)}{x_j+x_k}
    +\frac{s_{jk}(x_j-x_k)}{x_j+x_k}\biggr]^2\nn
    \\
    &\qquad\quad
    +\frac{s_{ijk}}{2s_{jk}}\biggl[\frac{x_i^3+1}{(1-x_i) x_i}
    +\frac{x_i (x_k-x_j)^2-2 (x_i+1)x_j x_k}{(1-x_i) x_i (1-\epsilon)}\biggr]
    +\frac{\epsilon }{2}-\frac{1}{4}\biggr\}+(2\leftrightarrow3)\ ,\nn
    \\[2ex]
    &P^{(0)}_{g\to ggg}(x_i,x_j,x_k;s_{ijk},s_{ij},s_{ik},s_{jk})
    \\
    &=
    C_A^2\biggl\{
    \frac{s_{ijk}^2}{s_{ij} s_{ik}}
    \biggl[\frac{2x_i(1\!+\!x_i)+1}{2(1\!-\!x_j)(1\!-\!x_k)}
    +\frac{1\!-\!2(1\!-\!x_i)x_i}{2x_j x_k}
    +\frac{x_i x_j(1\!-\!x_j)(1\!-\!2x_k)}{(1\!-\!x_k)x_k}
    +\frac{x_i}{2}(1\!+\!2x_i)\!+\!x_j x_k\!-\!2\biggr]\nn
    \\
    &\qquad\quad
    +\frac{s_{ijk}}{s_{ij}}\biggl[\frac{4 (x_i x_j-1)}{1-x_k}+\frac{x_i x_j-2}{x_k}
    +\frac{[1-x_k(1-x_k)]^2}{(1-x_i) x_i x_k}
    +\frac{5x_k}{2}+\frac{3}{2}\biggr]\nn
    \\
    &\qquad\quad
    +\frac{1}{4s_{ij}^2}(1-\epsilon) 
    \biggl[\frac{s_{ij} (x_i-x_j)}{x_i+x_j}
    +\frac{2(s_{jk} x_i-s_{ik} x_j)}{x_i+x_j}\biggr]^2
    +\frac{3 (1-\epsilon)}{4}\biggr\}+(\text{all permutations})\ .\nn
\end{align}

\bibliography{fjf_RG.bib}
\bibliographystyle{JHEP}

\end{document}